\documentstyle[floats,aps]{revtex}
\begin{document}
\draft
  \font\twelvemib=cmmib10 scaled 1200
  \font\elevenmib=cmmib10 scaled 1095
  \font\tenmib=cmmib10
  \font\eightmib=cmmib10 scaled 800
  \font\sixmib=cmmib10 scaled 667
  \skewchar\elevenmib='177
  \newfam\mibfam
  \def\mib{\fam\mibfam\tenmib}
  \textfont\mibfam=\tenmib
  \scriptfont\mibfam=\eightmib
  \scriptscriptfont\mibfam=\sixmib
  \mathchardef\alpha="710B
  \mathchardef\beta="710C
  \mathchardef\gamma="710D
  \mathchardef\delta="710E
  \mathchardef\epsilon="710F
  \mathchardef\zeta="7110
  \mathchardef\eta="7111
  \mathchardef\theta="7112
  \mathchardef\kappa="7114
  \mathchardef\lambda="7115
  \mathchardef\mu="7116
  \mathchardef\nu="7117
  \mathchardef\xi="7118
  \mathchardef\pi="7119
  \mathchardef\rho="711A
  \mathchardef\sigma="711B
  \mathchardef\tau="711C
  \mathchardef\phi="711E
  \mathchardef\chi="711F
  \mathchardef\psi="7120
  \mathchardef\omega="7121
  \mathchardef\varepsilon="7122
  \mathchardef\vartheta="7123
  \mathchardef\varrho="7125
  \mathchardef\varphi="7127

    \mathchardef\Gamma="7100
    \mathchardef\Delta="7101
    \mathchardef\Theta="7102
    \mathchardef\Lambda="7103
    \mathchardef\Xi="7104
    \mathchardef\Pi="7105
    \mathchardef\Sigma="7106
    \mathchardef\Upsilon="7107
    \mathchardef\Phi="7108
    \mathchardef\Psi="7109
    \mathchardef\Omega="710A

\def\frac#1#2{{\textstyle{#1 \over #2}}}
\def\half{\frac{1}{2}}
\def\fourth{\frac{1}{4}}
\def\nd{^{\vphantom{\dagger}}}
\def\yd{^\dagger}
\def\viz{{\it viz.\/}}
\def\etal{{\it et al.\/}}
\def\ie{{\it i.e.\/}}
\def\ket#1{{\,|\,#1\,\rangle\,}}
\def\bra#1{{\,\langle\,#1\,|\,}}
\def\braket#1#2{{\,\langle\,#1\,|\,#2\,\rangle\,}}
\def\expect#1#2#3{{\,\langle\,#1\,|\,#2\,|\,#3\,\rangle\,}}
\def\gtwid{\,{\raise.3ex\hbox{$>$\kern-.75em\lower1ex\hbox{$\sim$}}}\,}
\def\ltwid{\,{\raise.3ex\hbox{$<$\kern-.75em\lower1ex\hbox{$\sim$}}}\,}
\def\vph{\vphantom{\sum_i}}
\def\bvph{\vphantom{\sum_N^N}}
\def\sgn{\,{\rm sgn\,}}
\def\vphi{{\mib\varphi}}

\def\sbl{\left [}
\def\sbr{\right ]}
\def\({\left (}
\def\){\right )}
\def\cbl{\left\{}
\def\cbr{\right\}}

\def\cA{{\cal A}}
\def\cE{{\cal E}}
\def\cB{{\cal B}}
\def\cL{{\cal L}}
\def\cH{{\cal H}}
\def\cO{{\cal O}}
\def\cS{{\cal S}}

\def\Vx{{\vec x}}
\def\Vr{{\vec r}}
\def\Vx{{\vec x}}
\def\Ve{{\vec e}}
\def\VD{{\vec D}}
\def\Vv{{\vec v}}
\def\VX{{\vec X}}
\def\VP{{\vec P}}
\def\VQ{{\vec Q}}
\def\VR{{\vec R}}
\def\VF{{\vec F}}
\def\VA{{\vec A}}
\def\VE{{\vec E}}
\def\Vbeta{{\vec\beta}}
\def\VDelta{{\vec\Delta}}
\def\Veta{{\vec\eta}}
\def\VZ{{\vec Z}}
\def\VJ{{\vec J}}
\def\VB{{\vec B}}
\def\VV{{\vec V}}
\def\Vp{{\vec p}}
\def\Va{{\vec a}}
\def\Vq{{\vec q}}
\def\Vj{{\vec j}}
\def\VPi{{\vec\Pi}}
\def\VCA{\vec{\cal A}}
\def\Vk{{\vec k}}
\def\Vjm{{\vec\jmath}}
\def\Vxi{{\vec\xi}}

\def\rmz{{\rm z}}
\def\rmp{{\rm p}}
\def\rmv{{\rm v}}
\def\rmi{{\rm i}}
\def\rmC{{\rm C}}
\def\rmB{{\rm B}}
\def\rmH{{\rm H}}
\def\rmPsi{{\rm\Psi}}
\def\rms{{\rm s}}
\def\rme{{\rm e}}
\def\rmI{{\rm I}}
\def\rmOmega{{\rm\Omega}}
\def\rmTheta{{\rm\Theta}}

\def\eps{\epsilon}
\def\rhobar{{\bar\rho}}
\def\mnl{{\mu\nu\lambda}}
\def\abg{{\alpha\beta\gamma}}
\def\xhi{{\raise.35ex\hbox{$\chi$}}}
\def\pz{{\partial}}
\def\gmo{{\gamma-1\over\beta^2}}
\def\undertext#1{$\underline{\hbox{#1}}$}
\def\sss#1{{\scriptscriptstyle #1}}
\def\ss#1{{\scriptstyle #1}}
\def\ssr#1{{\sss{\rm #1}}}
\def\lala{\langle\!\langle}
\def\rara{\rangle\!\rangle}
\def\intl{\int\limits}
\def\ointl{\oint\limits}

\def\ehat{{\hat\rme}}
\def\qhat{{\hat q}}
\def\nhat{{\hat n}}
\def\Rhat{{\hat R}}
\def\vhat{{\hat v}}
\def\khat{{\hat k}}
\def\yhat{{\hat\rmz}}
\def\zhat{{\hat\rmz}}
\def\Dhat{{\hat\Delta}}
\def\betahat{{\hat\beta}}

\def\subhead#1{\leftline{\undertext{\bf{#1}}}\smallskip}
\def\subsubhead#1{\leftline{\undertext{\sl{#1}}}\smallskip}
\def\square{\vbox {\hrule height 0.6pt\hbox{\vrule width 0.6pt\hskip 3pt\vbox
{\vskip 6pt}\hskip 3pt \vrule width 0.6pt}\hrule height 0.6pt}}
\def\first{^\ssr{(1)}}
\def\lc{l_\circ}
\def\lt{\tilde l}
\def\ci{\,{\rm ci\,}}
\def\si{\,{\rm si\,}}
\def\grad{{\vec\nabla}}
\def\eff{{\rm eff}}
\def\QED{{\rm QED}}
\def\Rd{{\dot R}}
\def\Xd{{\dot X}}
\def\Yd{{\dot Y}}
\def\xd{{\dot x}}
\def\ydo{{\dot y}}
\def\VRd{{\dot \VR}}
\def\VXd{{\dot \VX}}
\def\VYd{{\dot \VY}}
\def\Vxd{{\dot \Vx}}
\def\Vyd{{\dot \Vy}}
\def\Vvs{{\Vv_\rms}}
\def\Vvsd{{{\dot \Vv}_\rms}}

\gdef\journal#1, #2, #3, 1#4#5#6{               
    {\sl #1~}{\bf #2}, #3 (1#4#5#6)}            
\def\nup{\journal Nucl. Phys., }
\def\jetp{\journal Sov. Phys. JETP, }
\def\jetpl{\journal Sov. Phys. JETP Lett., }
\def\jltp{\journal J. Low Temp. Phys., }
\def\ijmp{\journal Int. J. Mod. Phys., }
\def\prb{\journal Phys. Rev. B, }
\def\prl{\journal Phys. Rev. Lett., }

\twocolumn[\hsize\textwidth\columnwidth\hsize\csname @twocolumnfalse\endcsname

\title{Dynamical Vortices in Superfluid Films}
\author{Daniel P. Arovas and Jos\'e A. Freire}
\address{
Department of Physics, University of California at San Diego,
La Jolla, CA 92093}


\maketitle

\begin{abstract}
The coupling of superfluid film to a moving vortex is a gauge coupling
entirely dictated by topology.  From the definition of linking number,
one can define a gauge field $\cA^\mu$, whose (2+1)-dimensional curl is the
vortex 3-current $J^\mu$, and to which the superfluid is minimally
coupled.  We compute the superfluid density and current response to a
moving vortex.  Exploiting the analogy to $(2+1)$-dimensional
electrodynamics, we compute the effective vortex mass $M(\omega)$
and find that it is logarithmically divergent in the $\omega\to 0$ limit,
with a constant imaginary part, yielding a super-Ohmic dissipation in the
presence of an oscillating superflow.  Numerical integration of the
nonlinear Schr{\"o}dinger equation supports these conclusions.
The interaction of vortices with impurities coupling to the density also
is discussed.

\end{abstract}

\pacs{PACS numbers: 67.40.V}
\vskip2pc]

\narrowtext

\subhead{I. Introduction}

In this paper we investigate the effective action and dynamics of
vortices in compressible superfluid films at zero temperature.
In an {\it incompressible} two-dimensional (2D) superfluid, vortices behave
as massless charges in a uniform magnetic field -- their motion is along
an equipotential, the sum of logarithmic contributions from each of the
point ``charges'' \cite{don}.  The Lagrangian for a charge-neutral system
of vortices may be written
\begin{displaymath}
L=-\kappa\rhobar\sum_i n_i\,X_i\Yd_i + {\rhobar\,\kappa^2\over 2\pi}\!
\sum_{i<j} n_i n_j \ln|\VX_i-\VX_j|
\end{displaymath}
where $\kappa=h/m$ is the rotational quantum, $\rhobar$ is the bulk density
(or superfluid density \cite{rhos}), $n_i$ is the integer ``charge'' and
$\VX_i$ the position of the $i^\ssr{th}$ vortex.  The equations of motion,
\begin{displaymath}
\VXd_i= {\kappa\over 2\pi}\!\sum_{j\atop (j \neq i)} n_j
{\zhat\times(\VX_i-\VX_j)\over |\VX_i-\VX_j|^2}\ ,
\end{displaymath}
preserve the total potential energy of the vortices, which of course is
just the kinetic energy of the superfluid.  These equations are first order
in time -- there is no inertial term $\sum_i\half M_i \VXd{}_i^2$ in $L$.

In a compressible superfluid, the speed of sound $c$ is finite.  This leads
to retardation effects in the vortex dynamics.  Furthermore, accelerating
vortices may radiate phonons, leading to dissipation.  Both effects are
described by a complex frequency-dependent mass term $M(\omega)$, derived
below.  This physics is present in granular films and Josephson junction
arrays as well \cite{eck}.  The basic idea is to integrate out the phonons,
which represent a bosonic bath to which vortices are coupled, in the spirit
of Ref. \cite{CL}, and thereby derive an effective action for the vortices
alone.

This paper is organized as follows:
In section II we derive the analog of backflow for moving vortices in
superfluid films.  In section III we review the correspondence between
superfluid dynamics and electrodynamics in two space dimensions and show
how the results of section II may be obtained by a Lorentz transformation
of a static vortex solution.  Self-interaction effects, vortex mass, and
dissipation are discussed in section IV.  In section V, we report on the
results of numerical simulations of a vortex in an oscillating superflow,
from which we can extract $M(\omega)$ and compare with theoretical 
predictions.  Section VI discusses the interaction of vortices and
dynamical impurities.

\medskip

\subhead{II. Analog of Backflow for Dynamical Vortices}

Consider a vortex moving in a $(2+1)$-dimensional Bose fluid.  The only
information we have about the vortex is that it is a point object which
accrues a geometric phase of $2\pi$ in the many body boson propagator each
time it encircles a boson.  We write the vortex current density as \cite{fn1}
\begin{displaymath}
J^\mu=c\kappa\!\int\! d\tau\,\sum_l n_l {dX^\mu_l\over d\tau}\,
\delta^{(3)}(x-X_l(\tau))
\end{displaymath}
where $\tau$ parameterizes the vortex ``world lines'' $X^\mu_l(\tau)$,
which are one-dimensional filaments running through (2+1)-dimensional
spacetime.  The many boson Lagrangian is written 
\begin{displaymath}
\cL=\half m\sum_i \(d\Vx_i\over dt\)^2 -\sum_{i<j} v(|\Vx_i-\Vx_j|)
+\cL_{\rm top}
\end{displaymath}
where we assume a simple generic interacting Bose fluid (isotropic, single
component).  Here, $\cL_{\rm top}$ is the topological term in the Lagrangian
which counts the winding number of the vortices relative to the bosons.
This is explicitly written in terms of the linking number of their
trajectories \cite{fwaz},
\begin{eqnarray}
\cS_{\rm top}&=&\int\! d^2\!x\,dt\, \cL_{\rm top}=2\pi\hbar
\,N_{\rm link}\nonumber\\
&=& {1\over c}\!\int\! d^2\!x\,dt\,j^\mu{\eps_\mnl \pz^\nu\over
\pz^2} J^\lambda
\equiv\int\! d^2\!x\,dt\,j^\mu \cA_\mu
\label{action}
\end{eqnarray}
where the boson mass current density is $j^\mu=(c\rho,\Vj)$, and where
$\pz^\nu/\pz^2$ is a formal expression for a nonlocal operator.
Vortex current conservation, $\pz_\mu J^\mu=0$, allows one to construct
a gauge potential $\cA^\mu$ whose curl is $J^\mu$\cite{fwaz},
\begin{equation}
J_\mu=-c\,\eps_\mnl\,\pz^\nu\! \cA^\lambda\ ,
\label{Hopf}
\end{equation}
and thereby express the linking number as a local interaction between the
boson current density $j^\mu$ and the vortex gauge potential $\cA_\mu$.

The time-dependent Hamiltonian for the bosons in the presence of
moving vortices is thus
\begin{eqnarray}
\cH(A^\mu)&=&\cH(0)-\int\! d^2\!x\,j^\rmp_\mu(\Vx) \cA^\mu(\Vx,t)\nonumber\\
&&\qquad\qquad+\half\int\! d^2\!x\,\rho(\Vx)\,\VCA^{\,2}(\Vx,t)
\label{Ham}
\end{eqnarray}
where $\rho$ is the boson density and $\Vj^\rmp$ is given by
\begin{eqnarray*}
j_0^\rmp(x)&=&c\rho(\Vx)=mc\sum_i\delta(\Vx-\Vx_i)\\
\Vj^\rmp(\Vx)&=&\half\sum_i\sbl\,\Vp_i\,\delta(\Vx-\Vx_i)+
\delta(\Vx-\Vx_i)\,\Vp_i\,\sbr\ .
\end{eqnarray*}
The gauge invariant boson current density $j_\mu$ is then written \cite{bob}:
\begin{displaymath}
j_\mu=-{\delta\cH\over\delta \cA^\mu}=j_\mu^\rmp + \rho\, \cA_\mu
(1-\delta_{\mu 0})\ .
\end{displaymath}
At this point we have reproduced the well-known analogy between a rotating
superfluid and a superconductor in a magnetic field \cite{ajl}.  What is new
here is the explicit gauge-covariant, time-{\it dependent} description,
through Eqs. \ref{Hopf} and \ref{Ham}, of the
coupling of a superfluid to vortices, which are quanta of rotation.

The linear response of the boson system is given by the Kubo formula,
\begin{displaymath}
\langle j^\ssr{(1)}_\mu(\Vx,t)\rangle=
\int\! d^2\!x'\,dt'\,K_{\mu\nu}(\Vx,t;\Vx',t')\,\cA^\nu(\Vx',t')\ ,
\end{displaymath}
where
\begin{eqnarray*}
\lefteqn{K_{\mu\nu}(\Vx,t;\Vx',t')={i\over\hbar}\langle[j^\rmp_\mu(\Vx,t),
j^\rmp_\nu(\Vx',t')]\rangle\nd_0\,\rmTheta(t-t')}\\
&&\qquad-\langle\rho(\Vx,t)\rangle\nd_0\,
\delta_{\mu\nu}\,(1-\delta_{\mu 0})\,\delta(\Vx-\Vx')\,\delta(t-t')\ .
\end{eqnarray*}
The spatial part of $K_{\mu\nu}$ may be written in terms of longitudinal
and transverse components in Fourier space, \viz
\begin{displaymath}
K^{ij}(q)=-\qhat^i\qhat^j K\nd_\parallel(q)-
(g^{ij}+\qhat^i\qhat^j)
K\nd_\perp(q)\ .
\end{displaymath}
Gauge invariance, $K_{\mu\nu}(q) q^\nu=0$, may be used to relate the
$00$ and $0i$ components to $K_\parallel$:
\begin{displaymath}
K^{00}(q)=-{c^2|\Vq\,|^2\over\omega^2}\,K\nd_\parallel(q)\ \ ,\ \ 
K^{i0}(q)=-{c q^i\over\omega}\, K\nd_\parallel(q)\ .
\end{displaymath}
At zero temperature, the single mode approximation (SMA) gives for the
density response \cite{tql}
\begin{displaymath}
K^\ssr{SMA}_{00}(q)= {m c^2\rhobar\over\hbar}\,S(\Vq\,)\,\cbl
{1\over\omega + c|\Vq| + i\eps}-{1\over\omega - c|\Vq| + i\eps}\cbr\ ,
\end{displaymath}
where $\rhobar=\langle\rho\rangle$ is the average mass density and $S(\Vq\,)$
is the ground state static structure function.  Note that
$K^\ssr{SMA}_\perp(q)=\rhobar$, since the phonon is purely longitudinal.
Recall that $\lim_{\Vq\to 0} \hbar |\Vq\,|/2mc S(\Vq\,) = 1$ \cite{tql}.

\smallskip

{\it Static Vortex} -- We choose a gauge in which $\cA^0=0$ and
$\grad\!\times\!\VCA=n\kappa\,\delta(\Vx\,)\zhat$, which is satisfied by
$\VCA(\Vx,t)=n\kappa\,\zhat\times\Vx/2\pi|\Vx\,|^2$.  Now $\VCA(\Vq,\omega)=
-in\kappa\,(\zhat\times\Vq\,/|\Vq\,|^2)\cdot 2\pi\delta(\omega)$
is purely transverse, so the density response vanishes and the current density
response gives the usual $\langle\Vj^\ssr{(1)}(\Vx)\rangle=n\kappa\rhobar\,
\zhat\times\Vx/2\pi|\Vx\,|^2$.  The absence of a density variation in
response to the vortex seems to contradict the result that
$\delta\!\rho(r)/\rhobar=-n^2\,K_S\rhobar\,\hbar^2/2m^2|\Vx\,|^2$ far
from a vortex of strength $n$, where $K_S$ is the adiabatic
compressibility.  However, the $n^2$ dependence tells us that this is a
{\it nonlinear} response.  The second order response is formally written
\begin{displaymath}
\langle j_\alpha^\ssr{(2)}\rangle=\int\!d^2\!x' dt'\!\int\!d^2\!x'' dt''\,
R_\abg(x;x';x'') \cA^\beta(x') \cA^\gamma(x'').
\end{displaymath}
$R\nd_\abg\equiv R^\rmI_\abg + R^{\rm II}_\abg$
may be divided into two contributions.  The first, $R^\rmI_\abg$, is the
second order nonlinear susceptibility arising from the linear
$j_\mu^\rmp \cA^\mu$ coupling in $\cH$.  The second, $R^{\rm II}_\abg$, is the
linear susceptibility arising from the $\half\rho\VCA^{\,2}$
term -- this is given by the density-density correlation function, so
\begin{eqnarray*}
R^{\rm II}_\abg(x;x';x'')&=&-{1\over 2 c^2}K_{00}(\Vx,t;\Vx',t')\,
\delta(\Vx'-\Vx'')\\
&&\qquad\times\delta(t'-t'')\,\delta_{\beta\gamma}\,
(1-\delta_{\beta 0})\ .
\end{eqnarray*}
It is easy to check that the nonlinear response arising from $R^{\rm II}_\abg$
exactly reproduces the asymptotic density variation due to the vortex, and that
$R^{\rm I}_\abg$ does not contribute to $\cO(1/|\Vx\,|^2)$.

\smallskip

{\it Moving Vortex} -- Consider now a vortex moving with uniform velocity:
$\VX=\Vv t$.  We choose the gauge $\cA^0=0$ and
\begin{displaymath}
\VCA(\Vq,\omega)=-in\kappa\sbl {\zhat\times\Vq\over|\Vq\,|^2} +
{\Vq\over\omega|\Vq\,|^2}\,\zhat\cdot\Vv\times\Vq\,\sbr\cdot2\pi\,
\delta(\omega-\Vq\cdot\Vv\,)\ .
\end{displaymath}
Note that this is no longer purely transverse, so there will be a linear
response of the density to the moving vortex:
\begin{displaymath}
\langle\rho^\ssr{(1)}(\Vq,\omega)\rangle=8in\pi^2\rhobar\, S(\Vq\,)
{c\over |\Vq\,|} {\zhat\cdot\Vv\times\Vq\over (c\Vq\,)^2-(\Vv\cdot\Vq\,)^2}\,
\delta(\omega-\Vq\cdot\Vv\,)\ .
\end{displaymath}
Further assuming $|\Vv/c|\ll 1$, we obtain, at large distances
\begin{equation}
\langle\rho^\ssr{(1)}(\Vx,t)\rangle={n\kappa\rhobar\over 2\pi c^2}\,
{\zhat\cdot\Vv\times\Vx\over |\Vx-\Vv t|^2}\ ,
\label{denvar}
\end{equation}
which is identical to the result obtained by Duan \cite{duan2}.

The current density in the presence of a moving vortex is similarly computed
and found to be
\begin{displaymath}
\langle\Vj^\ssr{(1)}(\Vx,t)\rangle={n\kappa\rhobar\over 2\pi}\,
{\zhat\times\!\VR\over
\VR^2}\cbl 1 - {\Vv^{\,2}\over 2c^2}\sbl (\vhat\cdot\Rhat)^2
-(\vhat\times\Rhat)^2\sbr\cbr\ ,
\end{displaymath}
valid to order $\Vv^2/c^2$, with $\VR=\Vx-\Vv t$.

\medskip
\subsubhead{Contrast with Impurity Backflow}

It is important to contrast this behavior with the standard picture of
backflow in neutral systems.  A local perturbation coupling to the
superfluid as
\begin{displaymath}
\cH'=\int \! d^2\!x\>\rho(\Vx)\,U(\Vx,t)
\end{displaymath}
leads to a superfluid response \cite{tql}
\begin{eqnarray*}
\langle\rho\first(\Vq,\omega)\rangle&=&-{1\over c^2} K_{00}(\Vq,\omega)\,
U(\Vq,\omega)\\
\langle \Vj\first(\Vq,\omega)\rangle&=&-{1\over c} K^{i0}(\Vq,\omega)\,
U(\Vq,\omega)\,\ehat_i\ .
\end{eqnarray*}
Using the SMA response functions, and assuming $U(\Vx,t)=U_0\,
\delta(\Vx-\Vv t)$, one obtains
\begin{eqnarray}
\langle\rho\first(\Vx,t)\rangle&=&-{\rhobar\, U_0\over c^2}\,
\delta(\VR)\\
\langle \Vj\first(\Vx,t)\rangle&=&-{\rhobar\, U_0\over 2\pi c^2}\,
{v\over \VR^2}\cbl\vhat-2(\vhat\cdot\Rhat)\Rhat\cbr\ .\label{backflow}
\end{eqnarray}
The density response is purely local,
in contrast to that of Eq. \ref{denvar},
and the current, which vanishes in the static case $v=0$, is dipolar and
falls off as $1/R^2$.  The superfluid-vortex gauge coupling leads to a
much different linear response.

\medskip
\subhead{III. Analogy to QED$_{2+1}$}

These results may be understood in terms of the well-known correspondence
between ($2+1$)-dimensional superfluids and quantum
electrodynamics \cite{popov,flee,vol,sim,zee}, which we now
review.  One starts with the standard Ginzburg-Landau Lagrangian density
in the presence of an external gauge field $Z^\mu$,
\begin{eqnarray}
\cL[\Psi^*,\Psi]&=&\Psi^*(i\hbar\pz\nd_t+eZ^0)\Psi - {1\over 2m}\,
\Bigl|\({\hbar\over i}\grad+{e\over c}\VZ\)\Psi\Bigr|^2\nonumber\\
&&\qquad\qquad - \lambda \(|\Psi|^2-{\rhobar\over m}\)^2\ .
\label{GL}
\end{eqnarray}
For a superconductor, $Z^\mu$ would represent the electromagnetic
gauge potential, while in our case it can be used to describe an
externally imposed current.
At this point, the ``charge'' $e$ and velocity $c$ are arbitrary
parameters; we will take $c$ to be the speed of sound, defined below.
One substitutes $\Psi\equiv\sqrt{\rho/m}\,e^{i\theta} e^{i\xhi}$, where
$\theta(\Vx,t)$ is a smooth ``spin-wave'' field and $\xhi(\Vx,t)$ the
singular vortex field, which satisfies $J^\mu(x)=(\hbar c/m)\,\eps^\mnl
\pz\nd_\nu\pz\nd_\lambda\xhi$.  This gives
\begin{eqnarray*}
\cL'&=&-{\hbar\rho\over m}\(\pz\nd_t\theta+\pz\nd_t\xhi-{e\over\hbar}Z^0\)-
{\hbar^2\rho\over 2m^2}\(\grad\theta+\grad\xhi+{e\over\hbar c}\VZ\)^2\\
&&\qquad -{\hbar^2\over 8m^2\rho}\,(\grad\rho)^2-{\lambda\over m^2}
(\rho-\rhobar)^2
\end{eqnarray*}
after subtracting a time derivative term.  Decoupling the
$\(\grad\theta+\grad\xhi+e\VZ/\hbar c\)^2$ term, one arrives at
\begin{eqnarray*}
\cL''&=&
-{\hbar\VQ\over m}\cdot\(\grad\theta+\grad\xhi+{e\over\hbar c}\VZ\)
-{\hbar\rho\over m}\(\pz\nd_t\theta+\pz\nd_t\xhi-{e\over\hbar}Z^0\)\\
&&\qquad+{\VQ^2\over 2\rho}-{\hbar^2\over 8m^2}\,
{(\grad\rho)^2\over\rho}-{\lambda\over m^2}
\,(\rho-\rhobar)^2\ .
\end{eqnarray*}
Integrating over the spin wave field $\theta(\Vx,t)$ now generates the
constraint $\grad\cdot\VQ+\pz\nd_t\rho=0$ -- $\VQ$ is the mass current --
which is satisfied by introducing the gauge field $A^\mu=(A^0,\VA)$, where
\begin{displaymath}
(\rho-\rhobar,\VQ)=-{\rhobar\over c}\,
(\zhat\cdot\grad\!\times\!\VA,c\,\zhat\times\!\grad\! A^0+
\zhat\times\pz\nd_t\VA)\ .
\end{displaymath}
The coupling of the vortex current to this gauge field is due to the term
\begin{eqnarray*}
{\hbar\over m}\(\VQ\cdot\grad\xhi+\rho\,\pz\nd_t\xhi\)&=&{\hbar\rhobar\over m}
\,\eps^\mnl (A_\mu+a_\mu)\pz_\nu\pz_\lambda\xhi+ \pz(\ \cdot\ )\\
&=&{1\over c}\rhobar\,J^\mu\,(A_\mu+a_\mu)+ \pz(\ \cdot\ )\ ,
\end{eqnarray*}
where the {\it nondynamical} gauge field $a^\mu$ generates a static
magnetic field, \ie
\begin{displaymath}
\Ve=-\grad a^0-{1\over c}{\pz\Va\over\pz t}=0\quad,\quad
b=\zhat\cdot\grad\!\times\Va=-c\ ,
\end{displaymath}
which is satisfied by the gauge choice $\Va=\half\, c\, (y,-x)$, $a^0=0$,
for example.  The coupling between the gauge fields $A^\mu$ and $Z^\nu$
is then given by
\begin{displaymath}
{e\over m}\(\rho Z^0-{1\over c}\VQ\cdot\VZ\)={e\rhobar\over mc}\,
\eps^{\mu\nu\lambda} Z_\mu \pz_\nu(A_\lambda+a_\lambda)\ .
\end{displaymath}
Finally, one introduces the field strength tensor
\begin{displaymath}
F_{\mu\nu}=\pz_\mu A_\nu-\pz_\nu A_\mu =\pmatrix{0&E_x&E_y\cr
-E_x&0&-B\cr -E_y&B&0\cr}\ ,
\end{displaymath}
and obtains
\begin{eqnarray}
\cL_\eff&=&-{\rhobar\over 4}F_{\mu\nu} F^{\mu\nu}-{\rhobar\over c}
\,J^\mu(A_\mu+a_\mu)\nonumber\\
&&\qquad+{e\rhobar\over mc}\,\eps^{\mu\nu\lambda}
Z_\mu \pz_\nu (A_\lambda+a_\lambda)\nonumber\\
&&\qquad\qquad +{\rhobar\over 2c} {B\VE^2\over(1-B/c)}-
{\rhobar\over 8}\,{(\xi\grad B)^2\over(1-B/c)}\ ,
\label{fullqed}
\end{eqnarray}
where $c=\sqrt{2\lambda\rhobar/m^2}$ is the speed of sound and
$\xi=\hbar/mc$ is the coherence length.  Note
$K^\circ_\rms=m^2/2\rhobar^2\lambda$ is the bare compressibility.

The Lagrangian density $\cL_\eff$ describes ``charged'' particles
(vortices) moving in a background ``magnetic field'' $-c\,\zhat$,
minimally coupled to a dynamical gauge field $A^\mu$ (note that this is not
the gauge field defined in Eq. \ref{Hopf}).  That the background magnetic
field is the average boson density is of course due to the fact that the
vortices see the bosons as sources of geometric phase.  This was recognized
by Haldane and Wu\cite{berry}, who computed the Berry phase accrued by a
vortex as it executes adiabatic transport in the superfluid film.  If the
vortex position is $\Vxi$, and the adiabatic wavefunction is $\ket{\rmPsi}$,
then
\begin{displaymath}
\gamma\nd_\rmC=i\oint_\rmC d\Vxi\cdot\expect{\rmPsi}{\grad\nd_{\!\xi}}{\rmPsi}
=-2\pi{\rhobar\over m}\, S\nd_\rmC\ ,
\end{displaymath}
where $S\nd_\rmC$ is the area enclosed by the path $\rmC$ along which the
vortex travels.  Note that this immediately tells us that vortices
experience a Lorentz force when moving through a superfluid film.
In the nonrelativistic limit, their dynamics can be described
in terms of vortices being advected in each other's flow field, or
as charged particles in a background magnetic field moving under the
influence of each other's electric field.
A vortex-antivortex pair, for example, behaves like an exciton in a
magnetic field.

The linearized, long wavelength Lagrangian density is obtained by dropping
terms in $\cL_\eff$ which are higher than second order in the field
strength or which involve higher derivatives acting on the $A^\mu$ field.
One is then left with ($2+1$)-dimensional quantum electrodynamics,
\begin{displaymath}
\cL_\QED=-{\rhobar\over 4}\, F_{\mu\nu} F^{\mu\nu} +{\rhobar\over c}
\({e\over m}\eps^{\mu\nu\lambda}\pz_\nu Z_\lambda - J^\mu\)(A_\mu+a_\mu)\ ,
\end{displaymath}
in the presence of a uniform background magnetic field.  Note that the
gauge field $A^\mu$ couples to a sum of the (quantized) vortex current
density $J^\mu$ and the (not quantized) externally imposed ``current''
$(e/m)\eps^{\mu\nu\lambda}\pz_\mu Z_\lambda$, which could represent a
global rotation of the system \cite{ajl}.  When no vortices are present
($J^\mu=0$), one can integrate out the gauge field $A^\mu$ to obtain
\begin{eqnarray*}
S={\rhobar\over 2}\({e\over mc}\)^2\!\int\!\!{d^2\!k\,d\omega\over
(2\pi)^3}\,\Biggl[&& Z_\mu(-k)\(g^{\mu\nu}-{k^\mu k^\nu\over k^2}\) Z_\nu(k)\\
&&\quad+{e\rhobar\over m}\, Z^0(k)\,(2\pi)^3\,\delta^{(3)}(k)\Biggr] ,
\end{eqnarray*}
where $k^2=k^\mu k_\mu=c^{-2}\omega^2-\Vk^2$, and setting $e/mc\equiv 1$
one can read off the response tensor
\begin{displaymath}
K^{\mu\nu}(k)=\rhobar \(g^{\mu\nu}-{k^\mu k^\nu\over k^2}\)\ .
\end{displaymath}

When $Z^\mu=0$, one has a theory of vortices minimally coupled to the
gauge field $A^\mu$, and the action extremizing
equations for the fields are Maxwell's equations: $\pz_\mu F^{\mu\nu}=
J^\nu/c$, or
\begin{eqnarray*}
\grad\!\cdot\VE&=&{1\over c}J^0\\
\zhat\cdot\grad\!\times\VE&=&-{1\over c}{\pz B\over\pz t}\\
\grad B\times\zhat&=&{1\over c}\VJ+{1\over c}
{\pz\VE\over\pz t}\ .
\end{eqnarray*}
(Note $\grad\!\cdot\VB=\pz B/\pz z=0$, trivially.)
For the vortices, one has the Lorentz force law,
\begin{equation}
{d\VX\nd_l\over dX_l^0}={\zhat\times\VE(\VX\nd_l)\over
c - B(\VX\nd_l)}\ ,
\label{force}
\end{equation}
which says that the vortices move perpendicular to the local electric
field, with a magnetic field strength of $c-B$ which is the sum of
a uniform background contribution (the average boson density) and a
dynamical contribution (due to fluctuations in the boson density).

\medskip
\subsubhead{Lorentz Transformations}

To investigate the effects of moving vortices, it is useful to appeal to
the Lorentz invariance of $\cL_\QED$ and transform static solutions
\cite{broken}.
Recall that in ($2+1$)-dimensions, the Lorentz group has three generators,
corresponding to two boosts and one rotation.  The general boost transformation
is written
\begin{equation}
{L^\mu}_\nu=(gL)_{\mu\nu}=\pmatrix{\bvph\gamma&\gamma\beta_x&\gamma\beta_y\cr
\bvph\gamma\beta_x&\gmo\beta_x^2+1&\gmo\beta_x\beta_y\cr
\bvph\gamma\beta_y&\gmo\beta_x\beta_y&\gmo\beta_y^2+1\cr}
\label{boost}
\end{equation}
with $\Vbeta=\Vv/c$ and $\gamma=1/\sqrt{1-\beta^2}$.
Applying the Lorentz transformation $z'^\mu={L^\mu}_\nu z^\nu$ to the
coordinates and field strength tensor gives the familiar results
\begin{eqnarray*}
x^{\prime 0}&=&\gamma x^0+\gamma\Vbeta\cdot\Vx\\
\Vx'&=&\gamma x^0\Vbeta+\gmo(\Vbeta\cdot\Vx)\Vbeta+\Vx
\end{eqnarray*}
and
\begin{eqnarray*}
\VE'&=&\gamma\VE-\gmo(\Vbeta\cdot\VE)\Vbeta+\gamma B\zhat\times\Vbeta\\
B'&=&\gamma\zhat\cdot\Vbeta\times\VE+\gamma B\ .
\end{eqnarray*}
We may now transform solutions
\begin{displaymath}
\cbl x^\mu,J^\mu,F^{\mu\nu}\cbr\longrightarrow
\cbl x'^\mu,J'^\mu,F'^{\mu\nu}\cbr\ .
\end{displaymath}
A static charge $1$ vortex generates an electric
field $\VE=\kappa\Vx/2\pi|\Vx|^2$ and a magnetic field $B=0$.
Upon applying the boost of Eq. \ref{boost}, we obtain (dropping primes),
\begin{eqnarray}
\VE&=&{\gamma\kappa\over 2\pi}{x\nd_\perp \zhat\times\betahat
+ (x\nd_\parallel-\beta x^0)\betahat\over x_\perp^2+\gamma^2
(x\nd_\parallel-\beta x^0)^2}\nonumber\\
B&=&{\gamma\kappa\over 2\pi}{\beta x\nd_\perp\over x_\perp^2+\gamma^2
(x\nd_\parallel-\beta x^0)^2}
\label{eq6}
\end{eqnarray}
where we have written $\Vx=x\nd_\parallel\betahat+x\nd_\perp\zhat
\times\betahat$.

Now the rules for translating from $\VE$ and $B$ to $\Vj$ and $\rho$
are as follows:
\begin{displaymath}
\Vj=\rhobar\,\zhat\times\VE \qquad \rho=\rhobar\, (1-B/c)
\end{displaymath}
We now see that the linear response formulae (Eq. \ref{denvar} and accompanying
discussion) exactly reproduce these results to lowest order in $\beta$.
The vortex velocity is the the ratio $\Vj/\rho$, which is the content
of the Lorentz force law, Eq. \ref{force}.

\medskip
\subsubhead{A Tale of Two Vortices}

Consider now an elementary vortex-antivortex pair separated by a distance $a$.
We choose $\VX\nd_\pm(t)=vt\,\ehat_1\pm \half a\,\ehat_2$.
Computing the electric
and magnetic fields at one of the singularities due to the presence of the
other is easily accomplished with the Lorentz transformation.  One obtains
\begin{displaymath}
\VE(\VX\nd_\pm)=-{\gamma \kappa\over 2\pi a}\,\ehat_2\quad,\quad
B(\VX\nd_\pm)=-{\beta\gamma\kappa\over 2\pi a}\ .
\end{displaymath}
But if in the moving frame the vortices are stationary, we must have that
$c\Vbeta=d\VX\nd_\pm/dt$, which leads to the result
\begin{displaymath}
\beta(a)=\(\xi^2\over a^2+\xi^2\)^{1/2}\ .
\end{displaymath}
Thus, at large separations the pair's velocity is $c\,\xi/a$, but at smaller
separations the velocity asymptotically approaches the sound speed $c$.
We stress that this is true for the model defined by $\cL_\QED$,
where the vortices have no core.  The na{\"\i}ve expression $v(a)=c\,\xi/a$
begins to break down at distances on the order of $\xi$, where a proper
accounting of the terms neglected in the QED action must be taken in order
to reproduce the correct core structure.

\medskip
\subsubhead{Superflow and the Magnus Force}

In the above examples, we derived results for a moving vortex in a
stationary superfluid.  In this section, we make a Galilean transformation
(the original theory is Galilean invariant!) in order to discuss what
happens to vortices in the presence of a background superflow.
Starting with the Galilean-transformed Lagrangian density
\begin{displaymath}
\cL=i\hbar\,\Psi^*\pz\nd_t\Psi+i\hbar\Vv\cdot\Psi^*\grad\Psi-
{\hbar^2\over 2m}\,\bigl|
\grad\Psi\bigr|^2-\lambda\(|\Psi|^2-{\rhobar\over m}\)^2\ ,
\end{displaymath}
and proceeding as before, one derives the effective Lagrangian density
\begin{eqnarray*}
\cL_\eff&=&{\rhobar\over 2}\,\sbl{(\VE-B\zhat\times\Vbeta)^2\over 1-B/c}
-B^2\sbr-{\rhobar\over 8}\,{(\xi\grad B)^2\over 1-B/c}\\
&&\vph\qquad\qquad-{1\over c}\rhobar\, J^\mu(A_\mu+a_\mu)\ ,
\end{eqnarray*}
where $a^\mu$ now generates a static electric field as well as a static
magnetic field:
\begin{displaymath}
\Ve=-c\,\zhat\times\Vbeta\quad,\quad b=-c\ .
\end{displaymath}
The Lorentz force due to $\Ve$ is the Magnus force.  The vortex equation of
motion is
\begin{displaymath}
{d\VX\nd_l\over dX_l^0}=\Vbeta+{\zhat\times\VE(\VX\nd_l)-\Vbeta
B(\VX\nd_l)\over c - B(\VX\nd_l)}\ .
\end{displaymath}
When $\VE(\VX\nd_l)+\Ve=0$, the forces on vortex $l$ cancel, and it is
stationary.

\medskip
\subhead{IV. Self-Interaction, Inertial Mass, and Dissipation}

The analogy to electrodynamics suggests that there should be an electrodynamic
contribution to the mass and retardation effects,
as there are in ($3+1$)-dimensional classical electrodynamics\cite{jackson}.
In the superfluid, this is due to the phonon cloud carried by the vortex -- a
polaronic effect.  However, as we've seen, the coupling of vorticity to
superfluid density and current fluctuations is a gauge coupling which is rather
different from the local density coupling used in conventional polaron theories.
Still, this coupling is of the general form considered in Ref. \cite{CL},
\ie\ an external coordinate (the vortex position) coupled to a bath of
oscillators (the phonons).

We wish to integrate out the dynamical field $A^\mu$ corresponding to 
the phonon degrees of freedom and obtain an effective action for the
vortices.  Working in Lorentz gauge ($\pz_\mu A^\mu=0$), we integrate out
the Gaussian field $A^\mu$ in $\cL_\QED$ by solving the equations of
motion, yielding
\begin{displaymath}
A^\mu={1\over c}\,\square^{-1} J^\mu
\end{displaymath}
and an effective action for the vortices of
\begin{eqnarray*}
\cS_\eff&=&-{\rhobar\over c^3}\int\! d^3\!x\!\!\int\!d^3\!x'
J^\mu(x)\,\square^{-1}(x,x')\,J_\mu(x')\\
&&\qquad-{\rhobar\over c^2}\!\int\! d^3\!x\,J^\mu(x)\, a_\mu(x)\ .
\end{eqnarray*}
The inverse D'Alambertian has the form
\begin{displaymath}
\square^{-1}(x,x')={1\over 2\pi}{\rmTheta(x^0-x^{\prime 0}-|\Vx-\Vx'|)\over
\sqrt{(x^0-x^{\prime 0})^2-|\Vx-\Vx'|^2}}
\end{displaymath}
in ($2+1$)-dimensions and
\begin{displaymath}
\square^{-1}(x,x')={\delta(x^0-x^{\prime 0}-|\Vx-\Vx'|)\over
4\pi\,|\Vx-\Vx'|}
\end{displaymath}
in ($3+1$)-dimensions.  Thus, in contrast to the case of three spatial
dimensions, where $\square^{-1}$ vanishes unless $x-x'$ is light-like,
in our ($2+1$)-dimensional case
$\square^{-1}$ is nonzero everywhere inside the light cone\cite{evenodd}.
The finite sound speed $c$ leads to retardation effects.  One might
na{\"\i}vely think that this would lead to the collapse of the
vortex-antivortex pair, since the vortex should `see' the antivortex
at earlier times and {\it vice versa}.  However, although the {\it potentials}
are retarded, the {\it fields} of a uniformly moving charge point to the
{\it instantaneous} position of the charge (as we've derived above), and so
for the special case of a uniformly moving vortex-antivortex pair, there
is no apparent time delay\cite{sandy}.  

The self-interaction part of $\cS_\eff$ for a vortex of strength $n$ is
\begin{eqnarray}
\cS_{\rm self}&=&-{n^2\kappa^2\rhobar\over 4\pi c}\int_{-\infty}^\infty\!\!\!
du\int_0^\infty\!\!\!d\sigma\> \sbl 1-\Vbeta(u)\cdot\Vbeta(u+\sigma)\sbr
\nonumber\\
&&\qquad\qquad\times{\rmTheta(\sigma-|\VX(u+\sigma)-\VX(u)|)\over
\sqrt{\sigma^2-|\VX(u+\sigma)-\VX(u)|^2}}
\label{self}
\end{eqnarray}
where we've taken $x^0(u)=u$ as a parameterization of the vortex
world line, and $\Vbeta(u)=d\VX(u)/du$.  If the integrand in $S_{\rm self}$
were well-behaved and allowed an expansion of the form
\begin{displaymath}
\cS_{\rm self}=\int_{-\infty}^\infty\!\!\!\! dt\cbl -m_0 c^2 +
\half m_1 \Vv^{\,2}(t) + \ldots\cbr
\end{displaymath}
then we would associate a rest mass with $m_0$ and an inertial mass
with $m_1$.  The remaining terms, involving higher derivatives and
powers of the velocity, would be negligible in the Newtonian limit.
However, the integral over $\sigma$ in Eq. \ref{self} diverges logarithmically,
both for large and small $\sigma$.  The small $\sigma$ divergence is remedied
by a proper treatment of the core structure, which lies beyond the QED
approximation.  The large $\sigma$ divergence,
on the other hand, is real.  In the case of the parameter $m_0$, this is to be
expected, since we know the energy of a static vortex diverges logarithmically
with the size of the system, owing to the slow fall-off of the current
density $|\Vj|\propto 1/r$.  In the electrodynamic language, the energy
density, $\cE=\half\rhobar\,(\VE^2+B^2)$, dies off as $1/r^2$ in the vicinity of
a static vortex, yielding a logarithmic divergence when integrated over the
system.  In ($3+1$)-dimensional classical electrodynamics, by contrast,
the energy density dies off as $1/r^4$, and there is no infrared divergence.
Now we ask whether $m_1$ is finite.  The answer again is no.  Since
$\cS_{\rm self}$ is a Lorentz scalar, for constant $\Vbeta$ one has
\begin{displaymath}
\cS_{\rm self}(\Vbeta)=-\sqrt{1-\beta^2}\int_{-\infty}^\infty\!\!\!\!dt\,
m_0\, c^2
\end{displaymath}
which says that $m_1=m_0$.
As recently emphasized by Duan\cite{duan1,duan2}, this may be understood
in terms of the density variation $\langle\rho^\ssr{(1)}\rangle
\propto\zhat\cdot\Vv\times\Vx/|\Vx-\Vv t|^2$,
which produces a logarithmically infinite energy shift
\begin{displaymath}
\Delta E ={1\over 2 K_S\rhobar^2}\int\!d^2\!x\,
[\delta\!\rho(\Vx)]^2\ .
\end{displaymath}
Similarly, the total momentum $\VP$ of the moving vortex,
\begin{displaymath}
\VP^\ssr{(1)}(t)=m\int\!d^2\!x\>{\Vj(\Vx,t)\over\rho(\Vx,t)}\ ,
\end{displaymath}
diverges logarithmically \cite{duan2}.

\medskip
\subsubhead{Frequency-Dependent Inertial Mass}

In this section we compute the low frequency inertial mass of a single
vortex and find that it is frequency-dependent and logarithmically
divergent as $\omega\to 0$.  We start by expanding the self-interaction
contribution $\cS_{\rm self}$ for a single vortex \cite{scot}:
\begin{eqnarray*}
\cS_{\rm self}&=&-{n^2\kappa^2\rhobar\over 4\pi c}\int_{-\infty}^\infty\!\!\!
du\int_0^\infty\!\!\!d\sigma\>\sbl
1-\Vbeta(u)\cdot\Vbeta(u+\sigma)\sbr\times\\
&&\qquad\qquad\times
\sbl {1\over\sigma}+{|\VX(u+\sigma)-\VX(u)|^2\over 2\sigma^3}+\ldots\sbr\\
&=&\cS_{\rm static}+\half n^2\!\!\int_{-\infty}^\infty\!\!{d\omega\over2\pi}\,
M'(\omega)\,\omega^2\, |\VX(\omega)|^2+\ldots
\end{eqnarray*}
where $\cS_{\rm static}$ is action for a static vortex of strength $n$.
The quantity $M'(\omega)$ is found to be
\begin{eqnarray}
M'(\omega)&=&2\mu\int_\delta^\infty\!\!{ds\over s}\,\sbl \cos\omega s
-{1-\cos\omega s\over\omega^2 s^2}\sbr\label{mass}\\
&=&-\mu\cbl\ci(|\omega|\delta)+{1-\cos\omega\delta\over\omega^2\delta^2}
+{\sin\omega\delta\over\omega\delta}\cbr\ ,\nonumber
\end{eqnarray}
where $\mu\equiv\pi\xi^2\rhobar=\pi\hbar^2/2\lambda$ is the ``core
mass'' of the vortex \cite{duan2}, and $\ci(z)$ is the cosine integral
\cite{gradr}.  We have introduced an
ultraviolet temporal cutoff $\delta\approx\xi/c$ to regularize the $s$
integrals.  This crudely accounts for the core structure of the vortex
which lies beyond the approximation afforded by $\cL_\QED$.
The important point is that the {\it infrared} divergence of
$(\cS-\cS_{\rm static})$ is suppressed by the finite frequency $\omega$,
leading to a low frequency mass $n^2 M'(\omega)$ which diverges logarithmically
as $\omega\to 0$:
\begin{displaymath}
M'(\omega)=\mu\cbl -\ln(|\omega|\delta)-
(\rmC+\frac{3}{2})+\frac{11}{24}\omega^2\delta^2 +\ldots\cbr\ ,
\end{displaymath}
where $\rmC=0.577215\ldots$ is Euler's constant.

The effective Lagrangian $\cL_\eff$ does contain a term $-\frac{1}{8}\rhobar\,
(\xi\nabla B)^2$ which is dropped in the long wavelength effective theory but
is still quadratic in the field strengths.  (It also breaks Lorentz
invariance.)  Retaining this term, there are no ultraviolet divergences,
and the mass is
\begin{eqnarray}
M(\omega)&=&{\mu\over 2\sqrt{\Delta(\omega)}}\>\ln\({\sqrt{\Delta(\omega)}
+1\over\sqrt{\Delta(\omega)}-1}\)+
{i\pi\mu\,\sgn\omega\over 2\sqrt{\Delta(\omega)}}\label{vmass}\\
\bvph&\equiv&M'(\omega)+iM''(\omega)\nonumber\\
\Delta(\omega)&\equiv&1+{\omega^2\xi^2\over c^2}\ ,
\end{eqnarray}
where we have now included the imaginary part $M''(\omega)$.
The logarithmic divergence at small $\omega$ is still present, but in the
large $\omega$ limit we find that $M'(\omega)$ vanishes as $\omega^{-2}$
and $M''(\omega)$ as $\omega^{-1}$.
The Fourier transform of $M(\omega)$ is then causal:
\begin{eqnarray*}
M(t)&=&{\pi\mu c\over 2\xi}\sbl I_0(ct/\xi) - L_0(ct/\xi)\sbr\,\rmTheta(t)\\
&=&{\pi\mu c\over 2\xi}\sbl 1-\frac{2}{\pi}(ct/\xi)+\fourth(ct/\xi)^2+
\ldots\ \sbr\qquad(t\to 0)\\
&=&{\mu\over t}\sbl 1-(\xi/ct)^2 +\ldots\ \sbr\qquad(t\to\infty)\ .
\end{eqnarray*}
where $I_0(z)$ and $L_0(z)$ are modified Bessel and Struve functions,
respectively \cite{gradr}.
The logarithmic frequency dependence has previously been obtained by Eckern
and Schmid \cite{eck}, who investigated vortices in granular films, and by
Stamp, Chudnovsky, and Barbara \cite{SCB} in the context of magnetic domain
walls.

When several vortices are present, the effective vortex action
for low frequencies becomes (see the Appendix)
\begin{eqnarray}
\cS_{\rm eff}&=&\half\!\!\int\!{d\omega\over 2\pi}\,M(\omega)\,
\omega^2\, \bigg|\sum_i n_i\,\VX_i(\omega)\bigg|^2+\ldots\nonumber\\
&&\qquad+{\rhobar\,\kappa^2\over 4\pi}\int\!\!dt\,\,{\sum_{i\neq j}}'
n_i n_j\ln |\VX_i(t)-\VX_j(t)|\nonumber\\
&&\qquad\qquad-\kappa\,\rhobar\int\!\!dt\,\sum_i n_i\,X_i(t)\,\Yd_i(t)\ ,
\label{many}
\end{eqnarray}
where the first term arises from an expansion of $J^\mu\,\square^{-1} J_\mu$
in terms of the vortex coordinates themselves \cite{foot}, and where the
prime on the sum is a zero total vorticity restriction: $\sum_i n_i=0$.
Notice that the first (``kinetic'') term, discussed in the Appendix,
involves only the total dipole moment
operator $\VD(t)=\sum_i n_i\VX_i(t)$; this fact is intimately connected
with Galilean invariance and the stability of superflow in the absence
of disorder.  Consider, for example, an elementary ($n=\pm 1$)
vortex-antivortex pair.  Let $\VX\equiv\half(\VX_+ +\VX_-)$ be the
``center of mass'' (CM) coordinate and $\Vx\equiv(\VX_+ -\VX_-)$ be the
dipole moment.  The CM coordinate appears only in the Berry phase term
of the Lagrangian, which is
\begin{displaymath}
L_{\rmB}=-\kappa\,\rhobar\,(X_+ \Yd_+-X_-\Yd_-)
=-\kappa\,\rhobar\,(X\ydo-Y\xd)
\end{displaymath}
up to a total time derivative.  Thus, a path integral over the CM
coordinates generates a delta function at each time step, enforcing
$d\Vx(t)/dt=0$ always.

\medskip
\subsubhead{Relative Importance of Inertial Terms}

To investigate the importance of inertial terms relative to those arising
from the Lorentz force, we consider the response of an isolated $n=1$ vortex
to a time dependent field $\Ve(t)$, which might represent a sudden switching
on of a superflow which will accelerate the vortex \cite{eck} or an oscillating
superflow.  We find that the velocity $\VV(\omega)=-i\omega \VX(\omega)$
satisfies
\begin{equation}
\VV(\omega)=\sbl {1\over 1-r^2(\omega)}
\sbr\zhat\times\Ve(\omega)
+\sbl {ir(\omega)\over 1-r^2(\omega)}\sbr\Ve(\omega)
\label{motion}
\end{equation}
where the dimensionless function
\begin{displaymath}
r(\omega)\equiv{\omega M(\omega)\over\kappa\rhobar}
={\omega \xi\over 2c}\,{M(\omega)\over\mu}\ ,
\end{displaymath}
shown in Fig.~1, describes the inertial and frictional aspects of
the vortex's motion.

We now see that inertial effects will be relatively unimportant at
frequencies where the ``inertial parameter'' $r(\omega)$ is small.
Note that $r'(\omega)$ vanishes both for very low and very
high frequencies, as shown in Fig.~1.  Taking $c\sim 100$ m/s and
$\xi\sim 5$\AA, one obtains a characteristic frequency 
$\omega_0\equiv c/\xi\approx 10^{11}$ Hz.
At low frequencies, both real and imaginary components of $r(\omega)$
are small, and inertial effects are relatively unimportant.
From Eq. \ref{motion}, we find that an elementary vortex in an oscillating
superflow will move at a Hall angle $\theta_\rmH(\omega)=\tan^{-1}|r(\omega)|$
relative to the $\zhat\times\Ve$ direction.  The power dissipation per unit
frequency is given by
\begin{equation}
P(\omega)=\kappa\rhobar\,{\rm Re}\sbl \VV(\omega)\cdot\Ve^{\,*}(\omega)\sbr
\mathrel{\mathop=_{\omega\to 0}}\omega\,M''(\omega)\,|\Ve(\omega)|^2
\label{dissipation}
\end{equation}
which means $P(\omega)={\pi\over 2}\mu v_\rms^2\,|\omega|$ for a superfluid
velocity oscillation of amplitude $v_\rms$.

\medskip
\subhead{V. Numerical Simulation}

Since the predictions of the linearized theory are essentially classical,
we should expect to see the aforementioned effects of phonon radiation
by solving the non-linear Schr{\"o}dinger equation (NLSE),
\begin{displaymath}
i{\dot\psi}=-\half\grad^2\psi + (|\psi|^2-1)\psi
\end{displaymath}
where we now measure all distances in units of $\xi$ and times in units
of $\xi/c$, and $\psi$ itself in units of $\sqrt{\rhobar/m}$.

The NLSE was numerically integrated on a two-dimensional grid using a
operator splitting method \cite{numrec}.
To impose the background oscillating superflow the condensate was defined as 
\begin{displaymath}
\psi(\Vx,t)=e^{i\Vvs(t)\cdot\Vx}\vphi(\Vx ,t),
\end{displaymath}
where $\Vvs(t)$ is the chosen time dependent superflow
(in units of $c$) uniformly defined over the whole region, and $\vphi
(\Vx ,t)$ is, initially, a static vortex solution of the NLSE. With
the initial state representing a condensate with a vortex
{\it plus} a superflow given by $\Vvs(t=0)$, $\psi(\Vx,t)$ was evolved
according to the NLSE. The resulting equation for $\vphi(\Vx,t)$ is
\begin{equation}
i{\dot\vphi} =(\Vvsd\cdot\Vx)\vphi-\half(\grad+i\Vvs)^2\vphi
+(|\vphi|^2-1)\vphi\ .
\label{phieqn}
\end{equation}

When $\Vvs$ is constant in time, Galilean invariance means that a solution 
to \ref{phieqn} is given by
\begin{eqnarray*}
\vphi(\Vx,t)&=&\exp(-i\Vvs^2 t/2)\,f(\Vx-\Vvs t)\\
0&=&-\half\grad^2\!f + (|f|^2-1)f\ .
\end{eqnarray*}
In particular, if initially $f(\Vx)$ is a static vortex solution of
the NLSE, the time evolution of $\psi(\Vx,t)$ represents a vortex
being rigidly translated with velocity $\Vv_\rms $. The deviations from
this  ``massless'' behavior will become evident as $\Vvs$ becomes
time-dependent. 

In the method used the RHS of \ref{phieqn} is split into two parts which are
successively integrated, making the algorithm first-order accurate in
time \cite{numrec}:
\begin{eqnarray*}
i{\dot\vphi}&=&-\half\grad^2\!\vphi -i\Vvs\cdot\grad\vphi\qquad
\hbox{(1st step)}\\
i{\dot\vphi}&=&\Vvsd\cdot\Vx\,\vphi + \half \Vvs^2\,\vphi + (|\vphi|^2-1)\vphi
\qquad\hbox{(2nd step)}\ .
\end{eqnarray*}

The first step was integrated using the Crank-Nicholson method \cite{numrec},
which is unconditionally stable and second-order accurate in time; the
second step was integrated exactly. The time step was $0.01~\xi/c$ and 
the time-dependent velocity was along the longest dimension of the grid,
a channel whose spatial dimensions were $256\times 400$ points, with a
spacing equal to 0.1 or 0.05 $\xi$.
Along the edges we adopted von Neumann boundary
conditions for $\vphi(\Vx,t)$, which means that the superfluid
velocity computed from $\psi(\Vx,t)$ was $\Vvs(t)$ 
at the begining and at the end of the channel
(this also implies that the initial circular
vortex flow field had to be slightly distorted).

There are other ways of imposing the superflow, for instance using the 
boundary condition $\nhat \cdot \grad \psi = i \Vvs\psi$ for
the full condensate. The
influence of the edges however takes some time to reach the center where
the vortex is located, which can be specially inconvenient if one wants
$\Vvs$ to vary rapidly with time, another problem is that the
velocity field one obtains is not spacially uniform and does not lend
itself so easely to a comparison with the eletrodynamical theory. 

Fig. 2 shows the effect of a constant acceleration on an initially
uniform condensate. Here one sees a wake in the superfluid density
propagating at the speed of sound from the beginning of the channel
towards the end, and a counterwake propagating in the opposite 
direction. This consequence of the imposed accelerated flow takes a finite 
amount of time to reach the channel center, where the vortex was placed
in the subsequent simulations, and is responsible for an observed delay
in the vortex response.
This effect is related to the finite compressibility of the superfluid
and does no harm to the observation of vortex oscillations, it is in
fact what causes them. 

A more dangerous effect is the reflection of the
wake off the end of the channel. 
To avoid it, one has to restrict the observation time
to about $40~\xi/c$, for a channel length of $40~\xi$. This reflection
probably explains problems related to vortex shedding that were 
observed in some cases at the end of the observation period.

\medskip
\subsubhead{Results of the Simulations}

The equation of motion one gets  for the vortex position
in the presence of a time dependent background superflow is ($n=\pm 1$)
\begin{displaymath}
-i \omega M(\omega) \VV (\omega) = \pm\kappa\rhobar\>\zhat\times 
\sbl\Vvs(\omega)-\VV(\omega)\sbr .
\end{displaymath}
Thus, in the absence of the
inertial term, the vortex drifts with the superflow. $M(\omega)$ is the
Fourier transform of the causal kernel, as above.
These dynamics imply that for a monochromatic flow one should get a
response only at the driving frequency or at the resonance where
$\omega^2 M^2(\omega)=\kappa^2\rhobar^2$, which for a frequency-independent
mass corresponds to cyclotron oscillations.

We considered different forms of time dependent flow and compared
the observed trajectories with the equations of motion above.

To simulate an oscillating flow we took $v_\rms(t)$ to be
\begin{displaymath}
v\nd_\rms(t)=v^\circ_\rms\,\sin (2\pi t/T),$$
\end{displaymath}
where $v^\circ_\rms$ ranged from 0.1 to 0.3 $c$ and $T$ ranged from 2 to 30
$\xi/c$.
The trajectories we have obtained display a nearly periodic structure
with a characteristic frequency equal to the driving frequency. 
We obtained the frequency dependence of $M(\omega)$, as
implied by the model's equation of motion, by taking the Fourier
transform of the trajectory and reading its amplitude at the driving
frequency according to
\begin{equation}
M(\omega) = \mp i\kappa\rhobar\,{X(\omega)\over\omega Y(\omega)}\ .
\label{massnum}
\end{equation}
The results are shown in Fig. 3 together with the functional form
obtained from the linearized effective QED theory.

Since we did not get a perfect steady state response, as can be seen
in a typical trajectory as in Fig. 4, it was not very clear which region of
the data array to use in taking the Fourier transform. We chose to
ignore an initial structureless region, corresponding to the delay
mentioned above, and to take several Fourier transforms using time intervals
equal to an integer number of periods, all starting at the same
point. Hence, with the same trajectory, we obtained several values of
$M(\omega)$, which were averaged. The error bars in the figures 
correspond to this averaging process.

We obtained $M(\omega)$ for 3 amplitudes of the oscillatory velocity
field, namely 0.3, 0.2 and 0.1 $c$. The last 2 cases gave very 
similar values for the mass, except at lower frequencies, where the
vortex response was further from a steady state, and we could not
observe several periods of oscillations. Nonetheless, the expected
qualitative behavior was observed.

The higher amplitude case, $v^\circ_\rms=0.3 c$, posed more difficulties 
at low frequencies. Up to the lowest frequency we were able to get, one
sees a qualitatively different behavior, which may be due to the onset
of nonlinearities not visible in the other cases.   

Note that in the QED theory the real and imaginary parts of $M(\omega)$
obey a Kramers-Kronig relation. We could not check if such a relation
existed in the measured $M(\omega)$ because the points obtained were
too scattered to allow for a reliable fit.

We also tried a pulse form for $v_\rms(t)$,
\begin{equation}
v_\rms(t)=-\sqrt{2e}~v_{\rm{max}} \(\frac{t}{T}\) e^{-t^2/T^2}.$$
\label{pulse}
\end{equation}
This flow would produce a gaussian displacement along the channel
for a ``massless'' vortex. What is observed in Fig. 5 is a delayed
main peak in the parallel direction with an accompanying structure
in the perpendicular direction.  The real part $M'(\omega)$ is displayed
in Fig. 6.  Since the channel is finite in both width and length,
the zero frequency limit of $M(\omega)$ should cross over to a finite
value given by the system size.  In our system, with length 40 $\xi$,
values of $M'(\omega)$ greater than $\ln 40 \approx 3.7$ are difficult
to interpret.

\medskip
\subhead{VI. Impurities and Vortices}

We consider as a simple model of an impurity a point object which couples
linearly to the boson density, $\rho=\rhobar(1-B/c)$.
The Lagrangian describing the impurity is taken to be
\begin{displaymath}
L_{\rm imp}=\sum_a\half m_a\VRd_a\!\!\!{}^2
+{\rhobar\over mc}\!
\int\!\!d^2\!x\,\sum_a U_a(|\Vx-\VR_a|)\,B(\Vx)
\end{displaymath}
where $\VR_a(t)$ is the position of the $a^\ssr{th}$ impurity.
Upon integrating out the gauge field $A^\mu(\Vx,t)$, we obtain the
effective action
\begin{eqnarray}
\cS_{\rm eff}&&[\{\VX_i(t)\},\{\VR_a(t')\}]=\nonumber\\
&&\int\!\!dt\cbl-\half\kappa\,\rhobar\sum_i n_i\,\eps^{\alpha\beta}\,
X_i^\alpha\,\Xd_i^\beta
+\sum_a\half m_a\VRd_a\!\!\!{}^2\cbr\nonumber\\
&&+{\rhobar\,\kappa^2\over 4\pi}\int\!\!dt\,\,{\sum_{i\neq j}}'
n_i n_j\ln |\VX_i(t)-\VX_j(t)|\nonumber\\
&&-\int\!\!dt\!\!\int\!\!dt'\!\!\int\!\!{d^2\!k\over(2\pi)^2}
\!\!\int\!\!{d\omega\over 2\pi}\>{e^{i\omega(t-t')}\over \omega^2-\omega^2(k)}
\times\cbl\bvph\right.\nonumber\\
&&\left.\half\rhobar\,\kappa^2\sum_{i,j} n_i n_j\,
e^{-i\Vk\cdot(\VX_i(t)-\VX_j(t'))}\,\zhat\!\!\times\!\!\khat\!
\cdot\!\!\VXd_i(t)\,
\zhat\!\!\times\!\!\khat\!\cdot\!\!\VXd_j(t')\right.\nonumber\\
&&\left.-{i\rhobar\,\kappa\over m}\sum_{i,a} n_i\,|\Vk|\, U_a(k)\,
e^{-i\Vk\cdot(\VX_i(t)-\VR_a(t'))}\,\zhat\!\!\times\!\!\khat\!\cdot\!\!
\VXd_i(t)\right.
\nonumber\\
&&\left.+{\rhobar\over 2m^2}\sum_{a\neq b}\Vk^2\,U_a(k)\,U_b(-k)\,
e^{-i\Vk\cdot(\VR_a(t)-\VR_b(t'))}\cbr
\label{BIG}
\end{eqnarray}
where $\omega^2(k)=c^2 k^2 (1+\fourth\xi^2 k^2)$ is the dispersion relation
derived from the nonlinear Schr\"odinger equation.  (Superscripts in
Eq. \ref{BIG} refer only to spatial indices.)  We now focus on the
final two terms, which involve the impurity coordinates.

The last contribution in Eq. \ref{BIG} describes the interaction
between impurities mediated by phonons, including a self-interaction
term analogous to the self-interaction of vortices already discussed.
Ignoring retardation effects, the purely local superfluid response
(see Eq. \ref{backflow}) means that the impurity-impurity interaction
will be short-ranged provided the $U_a(k)$ are nonsingular in the infrared.
The self-interaction term contributes a mass shift
\begin{equation}
\Delta m_a(\omega)={\rhobar\over m^2}\int\!\!{d^2\!k\over (2\pi)^2}\,
{k^4\,|U_a(k)|^2\over \omega^2(k) \(\omega^2(k)-\omega^2\)}
\label{ishift}
\end{equation}
whose imaginary part (taking $\omega(k)=ck$),
\begin{displaymath}
\Delta m''_a(\omega)={\rhobar\over 4 m^2 c^6}\,|U_a(\omega/c)|^2\,\omega^2
\end{displaymath}
is super-Ohmic provided $U(k\to 0)\propto k^{-\sigma}$ with
$\sigma <\frac{3}{2}$ \cite{CL}.

Consider now the vortex-impurity interaction term.  We will ignore retardation
effects, and further assume a point interaction between impurities
and superfluid, so that $U_a(k)=U_a$ is a constant.  Without loss of generality,
we may consider a single vortex-impurity pair.  The contribution to the
effective action is then
\begin{eqnarray*}
\cS_{\rmv-\rmi}&=&-{nU\rhobar\,\kappa\over 2\pi mc^2}
\!\int\!\!dt\>{\zhat\!\times\!(\VX-\VR)\over (\VX-\VR)^2}\cdot\VXd\\
&=&-{nU\rhobar\,\kappa\over 2\pi mc^2}\!\int\!\!dt\,\VXd\!\cdot\!
{\pz\over\pz\VX}
\,\Theta(\VX-\VR)
\end{eqnarray*}
where $n$ is the integer charge of the vortex, $U$ is the impurity-boson density
coupling, and $\Theta(\Vx)=\tan^{-1}(y/x)$ is the angle function.  Note that if
$\VR(t)$ is time-independent, then, for closed paths,
$\cS_{\rmv-\rmi}$ is a topological quantity equal
to $-(nU\rhobar\,\kappa/mc^2)W_{\rmv-\rmi}$, where $W_{\rmv-\rmi}$
is the winding number of the vortex
about the impurity.  This makes excellent sense: the vortex effectively
counts the number of bosons it encircles, and the density response of the
superfluid generates a point accumulation of $\Delta N=-U\rhobar/m^2 c^2$
bosons to ``screen'' the impurity.

Now let the impurity move throughout the superfluid.  Varying
$\cS_{\rmv-\rmi}$ with respect to $X^\alpha(t)$, we find the
force on the vortex to be
\begin{displaymath}
F^\alpha={\delta\cS_{\rmv-\rmi}\over
\delta X^\alpha}=-{nU\rhobar\,\kappa\over 2\pi mc^2}\cbl{\delta^{\alpha\beta}
-2\Dhat^\alpha \Dhat^\beta\over\VDelta^2}\cbr\eps^{\beta\gamma}\Rd^\gamma
\end{displaymath}
with $\VDelta\equiv\VX-\VR$.  This is simply related to the backflow current.
If we ignore the self-interaction term for the vortices, which is appropriate
at very low frequencies, then the vortex equation of motion,
\begin{displaymath}
n\rhobar\,\kappa\,\eps^{\alpha\beta}\,\Xd^\beta=F^\alpha
\end{displaymath}
just says that the vortex moves in the dipolar backflow field of the moving
impurity.

Finally, consider the force on the impurity due to a moving vortex.
We find, neglecting the impurity mass renormalization term,
\begin{displaymath}
m_\ssr{imp} {\ddot R}^\alpha={nU\rhobar\,\kappa\over 2\pi mc^2}
\cbl{\delta^{\alpha\beta}-2\Dhat^\alpha \Dhat^\beta\over\VDelta^2}
\cbr\eps^{\beta\gamma}\Xd^\gamma\ .
\end{displaymath}
This too has a simple interpretation.  The quantity $U\rho(\VR)/m$ is the
local potential in which the impurity moves, and hence the force on the
impurity is $\VF=-(U/m)\grad\rho$.  Now recall Duan's result (Eq. \ref{denvar})
for the density response to a moving vortex.  Taking the gradient gives us
the appropriate force.

\medskip
\subsubhead{Polaron Model of a Quantum Vortex}

Niu, Ao, and Thouless [NAT] have investigated a model of a quantum vortex
coupled to superfluid density fluctuations \cite{nat}.  They describe the
vortex as a nonrelativistic particle of mass $M_\rme$ in a background uniform
magnetic field (corresponding to the average superfluid density) and assume a
{\it scalar} coupling to the phonons, \ie
\begin{eqnarray*}
\cH_\ssr{NAT}&=&{1\over 2M_\rme}\(\Vp+n\kappa\rhobar\,\Va/c\)^2 +
\sum_\Vk\hbar\omega\nd_\Vk (d\yd_\Vk d\nd_\Vk+\half)\\
&&\qquad+\sum_\Vk W(\Vk)\,e^{i\Vk\cdot\Vr}\,(d\nd_\Vk+d\yd_{-\Vk})\ ,
\end{eqnarray*}
where $n$ is the integer vorticity, $\grad\!\times\Va=-c\zhat$
accounts for the geometric phase due to the background bosons, and
$\omega\nd_\Vk$ is the phonon frequency at wavevector $\Vk$.
NAT go on to investigate a simple polaronic wavefunction which accounts for
the phonon cloud around a vortex and conclude that an infinite vortex mass 
would shrink the quantum uncertainty in the vortex position to zero, a
situation in conflict with explicit calculations using Feynman's trial 
vortex wavefunction.  At this level, the mass $M_\rme$ is phenomenological
and does not include the effects of phonons.  It may be perhaps more
appropriate to consider $M_\rme$ as the mass of an external particle
trapped in the vortex core, as considered by Demircan \etal\ \cite{dan}.

In keeping with the general philosophy that the vortex is a topological object
which couples to the boson density as in Eq. \ref{Ham}, we propose a variant
of the NAT model:
\begin{displaymath}
\cH={1\over 2M_\rme}\(\Vp+ n\kappa\rhobar\,(\Va+\VA)/c\)^2
+ \sum_\Vk\hbar\omega\nd_\Vk (d\yd_\Vk d\nd_\Vk+\half)\ ,
\end{displaymath}
where $\omega\nd_\Vk=\hbar ck$ and
\begin{displaymath}
\VA(\Vr)={i\over\sqrt{\rmOmega}}\sum_\Vk\sqrt{\hbar c\over 2\rhobar\,\bigl|\Vk
\bigr|}\,\zhat\times\khat\,e^{i\Vk\cdot\Vr}\,(d\nd_\Vk+d\yd_{-\Vk})
\end{displaymath}
is the quantized radiation field corresponding to the superfluid density
and current fluctuations \cite{baym} ($\rmOmega$ is the area of the
system).  We rewrite $\cH$ as
\begin{displaymath}
\cH={\VPi^2\over 2 M_\rme}+{n\kappa\rhobar\over M_\rme c}\,\VA\cdot\VPi+
{n^2\kappa^2\rhobar^2\over 2M_\rme c^2}\,\VA^{\,2}
\end{displaymath}
where $\VPi=\Vp-\half n\kappa\rhobar\,\zhat\times\Vr$ is the cyclotron
momentum operator for the vortex.  Note that $[\Pi_x,\Pi_y]=i\hbar^2/\lc^2$,
where $\lc\equiv\sqrt{m/2\pi n\rhobar}$ is the ``magnetic length''
for a vortex of strength $n$.  We now work to lowest order in $\VA$, and
follow NAT by assuming a trial state
\begin{displaymath}
\ket{\rmPsi[\VR,\cbl n_\Vk\cbr]}=\ket{\xhi\nd_\VR}\otimes
\ket{\rmPsi_{\rm ph}}
\end{displaymath}
with
\begin{displaymath}
\xhi\nd_\VR(\Vr)={1\over\sqrt{2\pi l^2}}\,e^{-(\Vr-\VR)^2/4l^2}\,
e^{-i\zhat\cdot\Vr\times\VR/2\lc^2}\ ,
\end{displaymath}
treating $l$ as a variational parameter.  Taking the expectation value of
$\cH$ is the state $\xhi\nd_\VR(\Vr)$, we obtain the effective phonon
Hamiltonian
\begin{eqnarray*}
\cH_{\rm ph}&=&\sum_\Vk\hbar\omega\nd_\Vk(d\yd_\Vk d\nd_\Vk+\half)
+\sum_\Vk W(\Vk)\,e^{i\Vk\cdot\VR}\,(d\nd_\Vk+d\yd_{-\Vk})\\
W(\Vk)&=&{n\kappa^2\over 4\pi c} {l^2\over\lc^2}{m\over M_\rme}
\sqrt{\rhobar\,\hbar\omega_\Vk\over 2\rmOmega}\,e^{-\half\Vk^2 l^2}\ .
\end{eqnarray*}
The phonon ground state is a coherent state,
\begin{displaymath}
\ket{\rmPsi_{\rm ph}}=\exp\cbl \sum_\Vk {W(\Vk)\over\hbar\omega\nd_\Vk}\,
(d\nd_\Vk\, e^{i\Vk\cdot\VR}-d\yd_\Vk\, e^{-i\Vk\cdot\VR})\cbr\ket{0}\ ,
\end{displaymath}
and the total energy is
\begin{eqnarray*}
E&=&\int\!d^2\!r\,\xhi^*_\VR(\Vr)\,\({\VPi^2\over 2M_\rme}\)\,
\xhi^{\vphantom{*}}_\VR(\Vr)-\sum_\Vk{\bigl|W(\Vk)\bigr|^2\over\hbar
\omega\nd_\Vk}\\
&=&{\hbar^2\over 4M_\rme\lc^2}\sbl {\lc^2\over l^2}+
(1-n^2\mu/M_\rme){l^2\over \lc^2}\sbr
\end{eqnarray*}
where $\mu=\pi\rhobar\,\xi^2=\kappa^2\rhobar/4\pi c^2$ as before.
Setting $\pz E/\pz l=0$ gives
\begin{displaymath}
l=\lc/\root 4 \of {1-n^2\mu/M_\rme}\ .
\end{displaymath}
Note that no solution exists for $M_\rme<n^2\mu$, which we interpret in the
following manner.  A cyclotron mode will show up as a pole in the denominator of
Eq. \ref{motion}, which means $\omega M_\rme(\omega)=\pm \kappa\rhobar$.
Now no such pole exists in the absence of the external mass $M_\rme$, but if
$M_\rme$ is added to our $M(\omega)$, then a damped cyclotron resonance does
exist at the cyclotron frequency $\omega\approx\kappa\rhobar/M_\rme$, provided
that $M_\rme\gtwid n^2\mu$.

NAT compute a renormalized magnetic length $\lt$ according to the relation
\begin{displaymath}
\Bigl|\braket{\rmPsi(\VR)}{\rmPsi(\VR+\Veta\,)}\Bigr|^2\equiv\exp\sbl-
{|\Veta\,|^2\over
2\lt^2}+\cO(|\Veta\,|^4)\sbr\ ,
\end{displaymath}
which is obtained from the overlaps
\begin{displaymath}
\braket{\xhi\nd_\VR}{\xhi\nd_{\VR+\Veta}}=
\exp\sbl -{i\zhat\cdot\VR\times\Veta
\over 2\lc^2}-{|\Veta\,|^2\over 8l^2}-{|\Veta\,|^2\over 8\lc^2}\sbr
\end{displaymath}
and
\begin{eqnarray*}
&&\braket{\rmPsi_{\rm ph}(\VR)}{\rmPsi_{\rm ph}(\VR+\Veta\,)}=\\
&&\qquad\qquad\exp\(-\sum_\Vk{\bigl|W(\Vk)\bigr|^2\over [\hbar\omega\nd_\Vk]^2}
(1-\cos\Vk\cdot\Veta\,)\)\ .
\end{eqnarray*}
We find
\begin{displaymath}
{1\over \lt^2}={1\over 2l^2}+\sbl 1+{\pi n\over 2\sqrt{2}}
\({m\over M_\rme}\)^2 {\xi^3 l\over \lc^4}\sbr {1\over 2\lc^2}\ .
\end{displaymath}

\medskip
\subhead{VII. Conclusion}

In this paper we have explored the theory of dynamical vortices in
superfluid films, deriving a frequency-dependent vortex mass which
enters into the vortex equations of motion, as well as describing
dissipation by radiation of phonons.  Numerical simulations corroborating
the predicted behavior of $M(\omega)$ were presented as well.
These calculations may be extended to $(3+1)$-dimensional superfluids
as well \cite{threed}.  These results will be presented in a future
publication.

{\it Acknowledgements} --- DPA gratefully acknowledges conversations with 
E. Demircan, J. Duan, A. Fetter, S. Kivelson, J.-M. Leinaas, L. Myklebust,
Q. Niu, S. Sondhi, P. C. E. Stamp, A. Stern, S.-C. Zhang, and particularly
S. R. Renn.  DPA and JAF are also grateful to H. Levine for his comments
and suggestions regarding the numerical simulations.
JF was supported in part by the Brazilian Government Agency CAPES.

\medskip
\subhead{Appendix: Inertial Term for Many Vortices}

The velocity-dependent part of the effective action induced by integrating
out the phonon field is
\begin{eqnarray*}
\Delta\cS&=&-\half\rhobar\,\kappa^2
\int\!\!dt\!\!\int\!\!dt'\!\!\int\!\!{d^2\!k\over(2\pi)^2}
\!\!\int\!\!{d\omega\over 2\pi}\>{e^{i\omega(t-t')}\over
\omega^2-\omega^2(k)}\cr
&&\bvph\times\sum_{i,j} n_i n_j\,
e^{-i\Vk\cdot(\VX_i(t)-\VX_j(t'))}\,\zhat\!\!\times\!\!\khat\!
\cdot\!\!\VXd_i(t)\,\zhat\!\!\times\!\!\khat\!\cdot\!\!\VXd_j(t')\ .
\end{eqnarray*}
We define $\VDelta_{ij}(t,t')\equiv\VX_i(t)-\VX_j(t')$.  Using
\begin{eqnarray*}
\int{d\khat\over2\pi}\,e^{-i\Vk\cdot\VDelta}\,\khat^\alpha\,\khat^\beta
&=&\half J_0(k\Delta)\delta^{\alpha\beta}\\
&&\quad+\half J_2(k\Delta)(\delta^{\alpha\beta}-2\Dhat^\alpha\Dhat^\beta)
\end{eqnarray*}
where $J_n(z)$ is the Bessel function of order $n$.  Taking $\omega(k)=ck$,
we have
\begin{eqnarray*}
\Delta\cS&=&-{\rhobar\,\kappa^2\over 8\pi c^2}
\int\!\!dt\!\!\int\!\!dt'\!\!\int\!\!{d\omega\over 2\pi}\,e^{i\omega(t-t')}
\sum_{i,j}n_i n_j\,\Xd_i^\alpha(t)\Xd_j^\beta(t')\\
&&\times\sbl\vph\(K_2(-i\omega\Delta_{ij}/c)+{2c^2\over\omega^2\Delta_{ij}^2}\)
(\delta^{\mu\nu}-2\Dhat^\mu\Dhat^\nu)\right.\\
&&\qquad\left.\vph-
K_0(-i\omega\Delta_{ij}/c)\delta^{\mu\nu}\sbr\eps^{\alpha\mu}\eps^{\beta\nu}
\end{eqnarray*}
where $K_n(z)$ is a modified Bessel function, and where
$\omega\to\omega + i 0^+$ is understood.
Since the above integrand is already quadratic in velocities, which we assume
are small compared with $c$, we may approximate $\Delta_{ij}$ as a constant.
Expansions of $K_n(z)$ for small $z$\cite{gradr} yield
\begin{eqnarray*}
K_0(-iz)&=&-\rmC-\ln(-iz/2)+\ldots\\
K_2(-iz)+{2\over z^2}&=&\frac{1}{8}z^2\ln(-iz/2)+\ldots
\end{eqnarray*}
and at low frequencies the first of these terms dominates, so provided
$\omega\ll c/\Delta_{\rm rms}$ we recover the action of Eq. \ref{many}.

\undertext{FIGURE CAPTIONS}

\vspace{0.1 in}

FIG.1 \quad Complex frequency-dependent vortex mass $M(\omega)$ from
eq. \ref{vmass}.  Also shown is the complex dimensionless inertial
parameter $r(\omega)$.

\smallskip

FIG.2 \quad Amplitude of $\phi(\Vx,t)$ for a uniformly accelerated condensate,
plotted as a function of length along the direction of flow, at ten equally
spaced time intervals separated by $\Delta t=5\xi/c$.

\smallskip

FIG.3 \quad Complex mass $M(\omega)$ inferred from eq. \ref{massnum}.
Solid curves are $M'(\omega)$ and $M''(\omega)$ from the linearized
electrodynamic theory.
(a) $M'(\omega)$ for driving amplitude $0.1\,c$  (b) $M''(\omega)$ for
driving amplitude $0.1\,c$.
(c) $M'(\omega)$ for driving amplitude $0.2\,c$.  (d) $M''(\omega)$ for
driving amplitude $0.2\,c$.
(e) $M'(\omega)$ for driving amplitude $0.3\,c$.  (f) $M''(\omega)$ for
driving amplitude $0.3\,c$.

\smallskip

FIG.4 \quad Typical vortex trajectory $\VX(t)$ with harmonic forcing of
period $T=5\xi/c$.

\smallskip

FIG.5 \quad Vortex trajectory $\VX(t)$ for a pulsed superflow given by
eq. \ref{pulse}.  In this case $T=2\xi/c$ and $v_{\rm max}=0.4c$.
The smooth curve is the pulse shape.

\smallskip

FIG.6 \quad Real part of the mass $M'(\omega)$ obtained from analysis of
response to the pulse flow.  Solid curve is the prediction of the
linearized electrodynamic theory.


\begin{references}
\bibitem{rhos} At zero temperature, in the absence of disorder,
the bulk density $\rhobar$ is the superfluid density $\rho_\rms$.
\bibitem{fn1}
Throughout this paper we write $x^\mu=(ct,x,y)$ and $q^\mu=(\omega/c,\Vq\,)$,
where $c$ is the speed of sound in the superfluid, and
$g_{\mu\nu}=\rm{diag}(+,-,-)$ is used to raise and lower indices.
\bibitem{sandy} DPA is indebted to A. Fetter for bringing this to his
attention.
\bibitem{broken} Strictly speaking, $\cL_\QED$ is not Lorentz-invariant,
owing to the nondynamical gauge field $a^\mu$.  However, this only breaks
the Lorentz invariance at the level of the equations of motion for the
vortices, where $c$ appears as a static magnetic field, and not
for the fields $\VE$ and $B$, which do obey Maxwell's equations.
\bibitem{foot} Note that the effective action is real since
$M''(\omega)$ is odd in frequency.
\end{references}
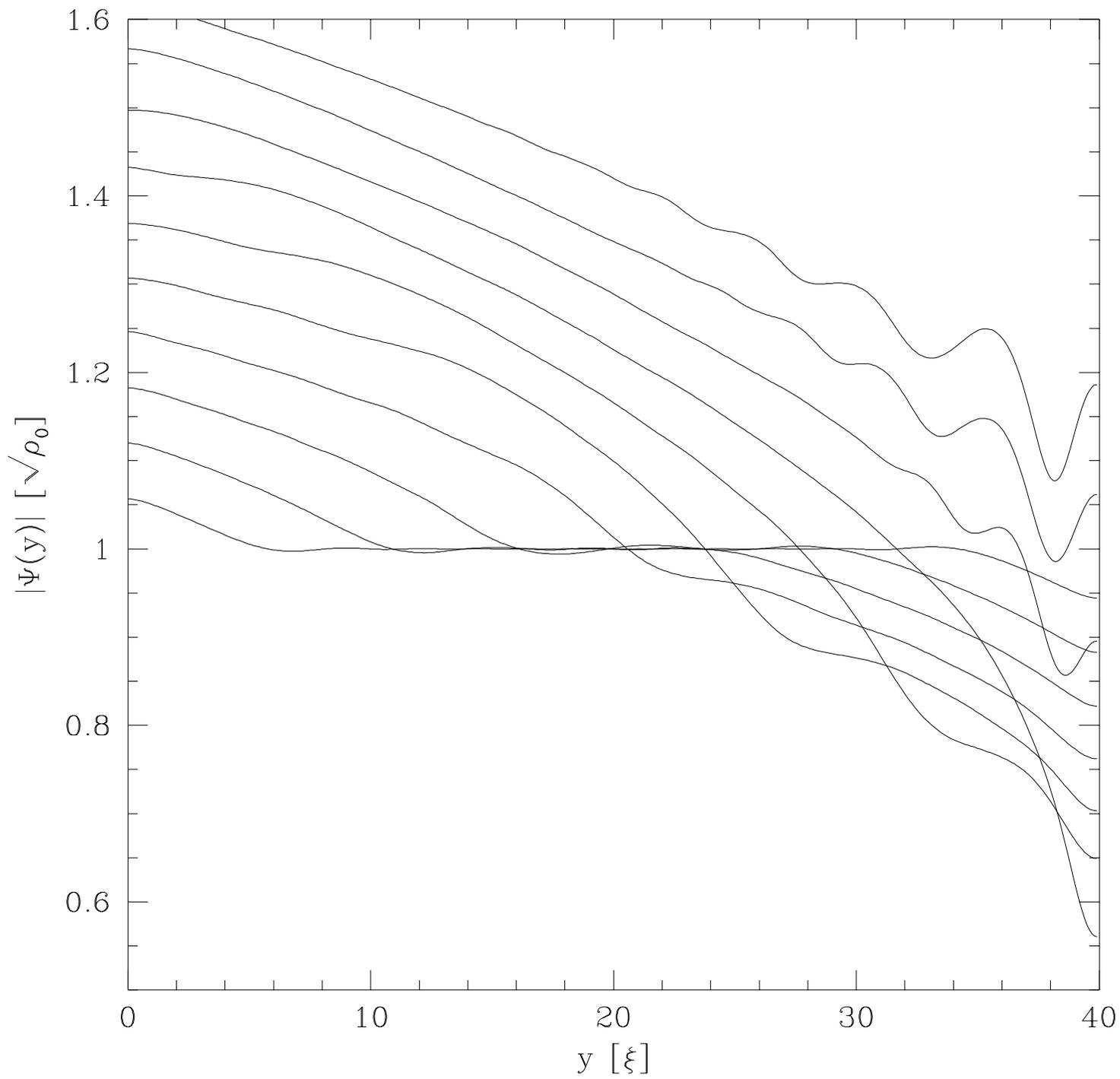
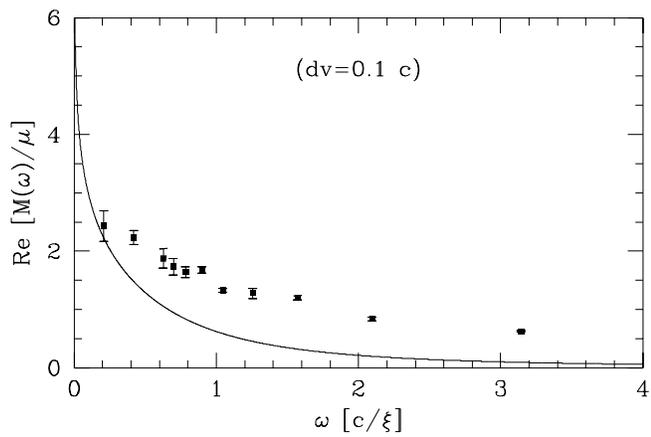
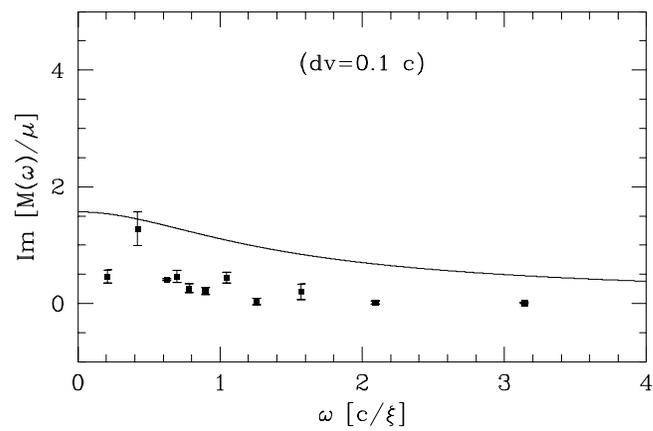
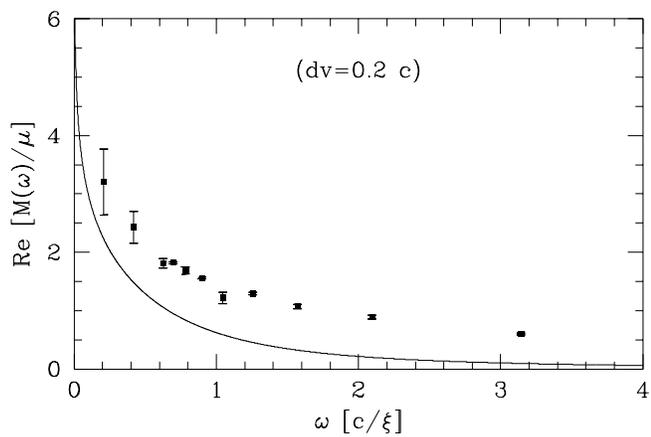
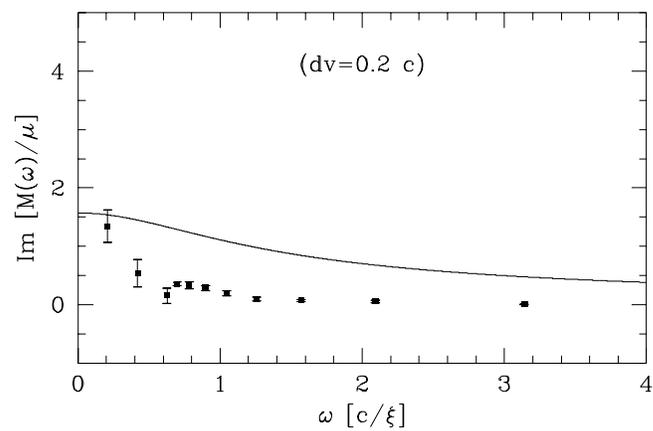
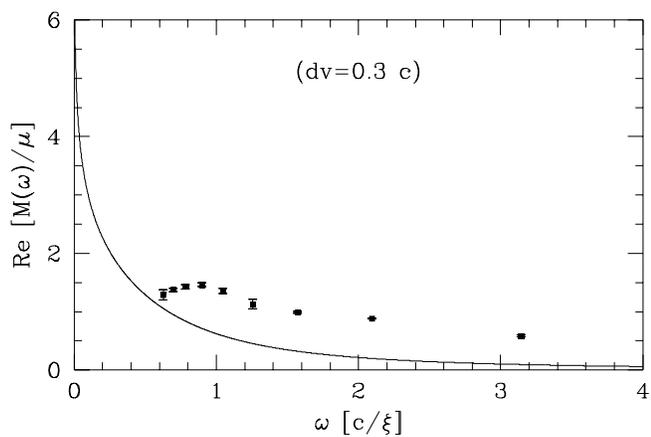
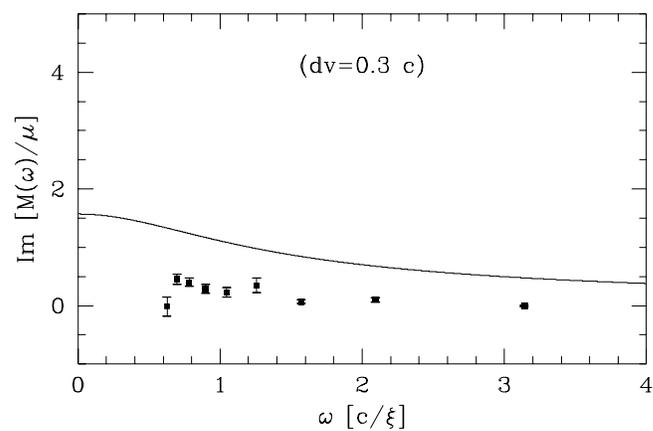
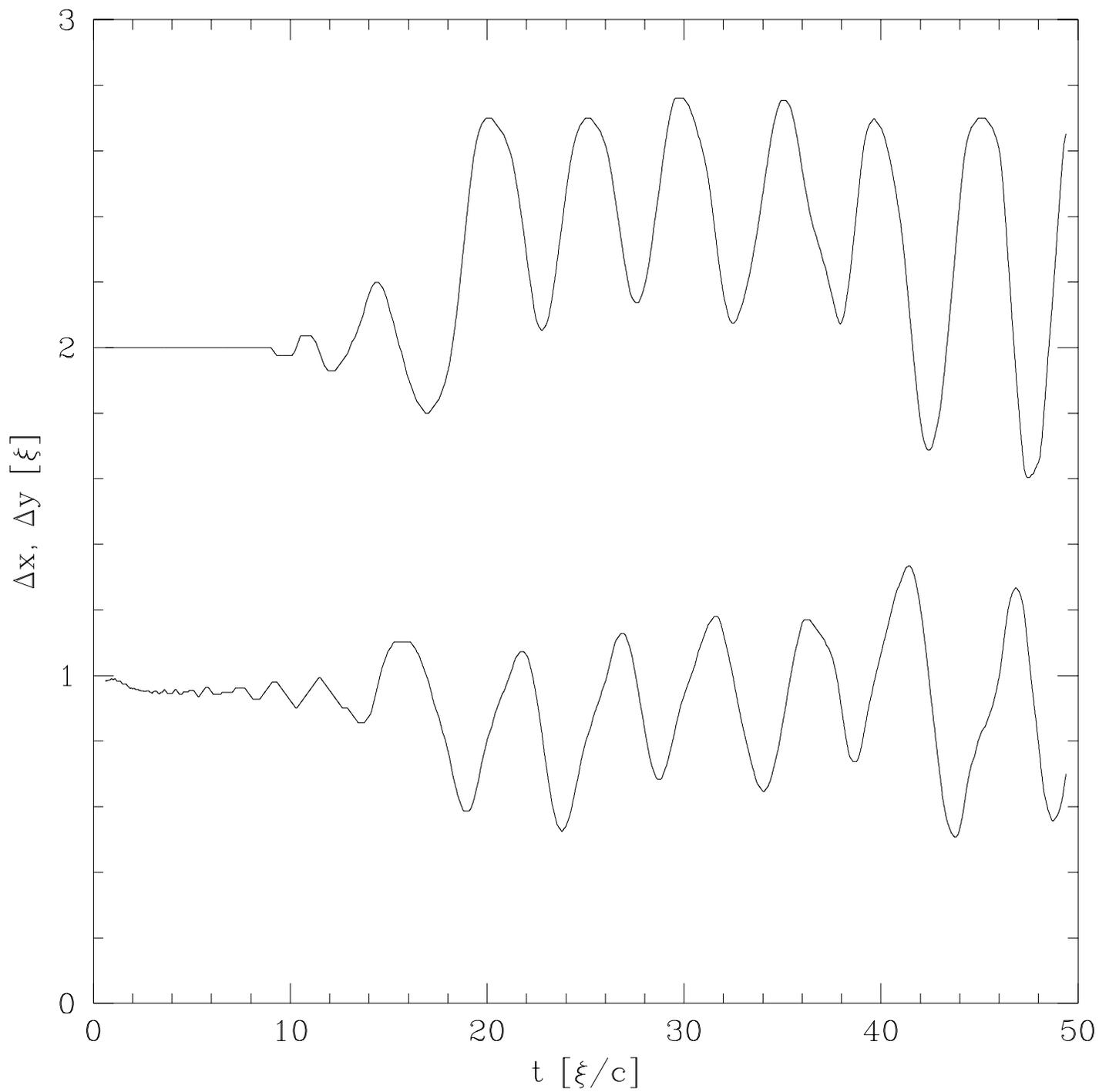
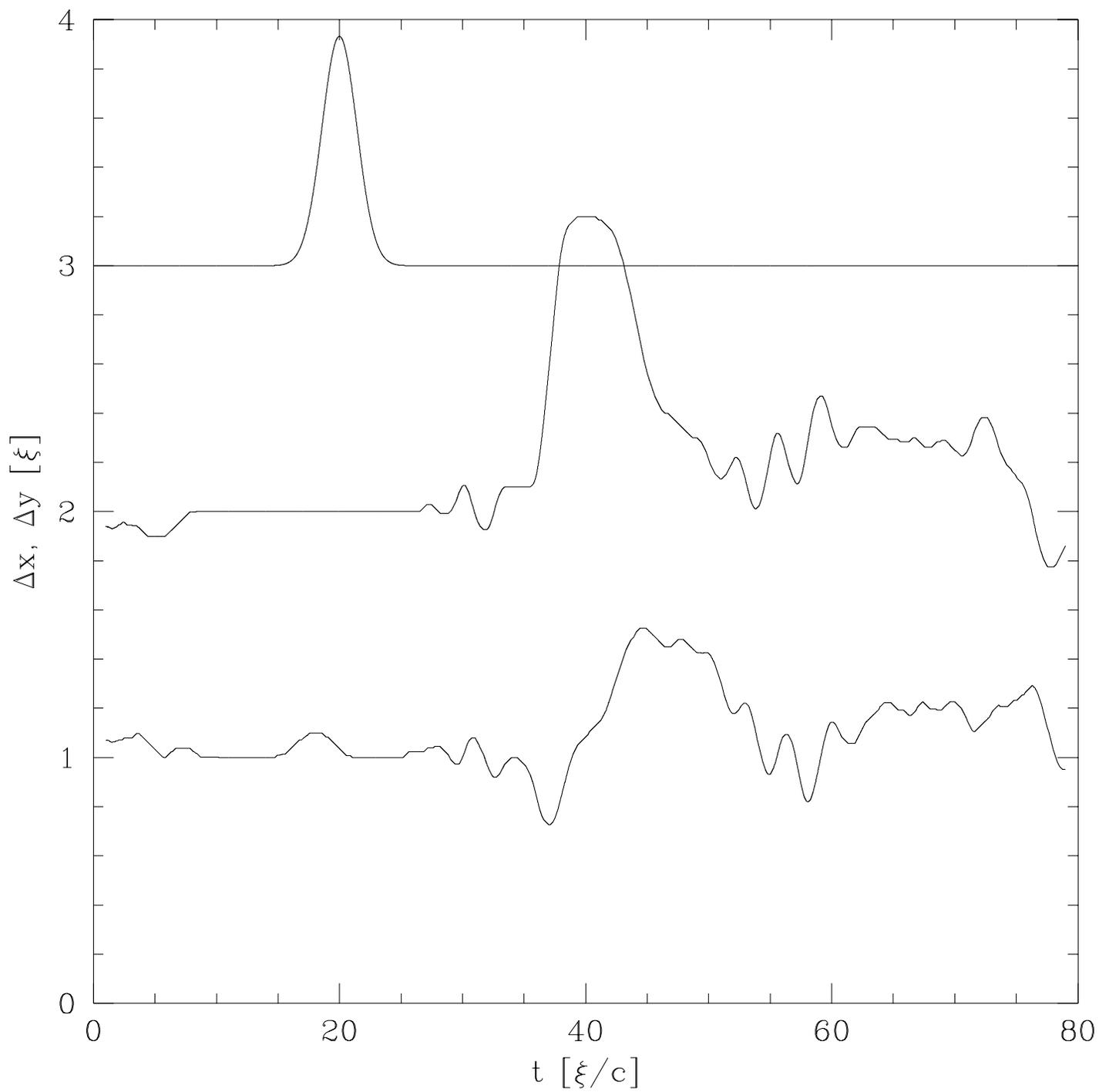
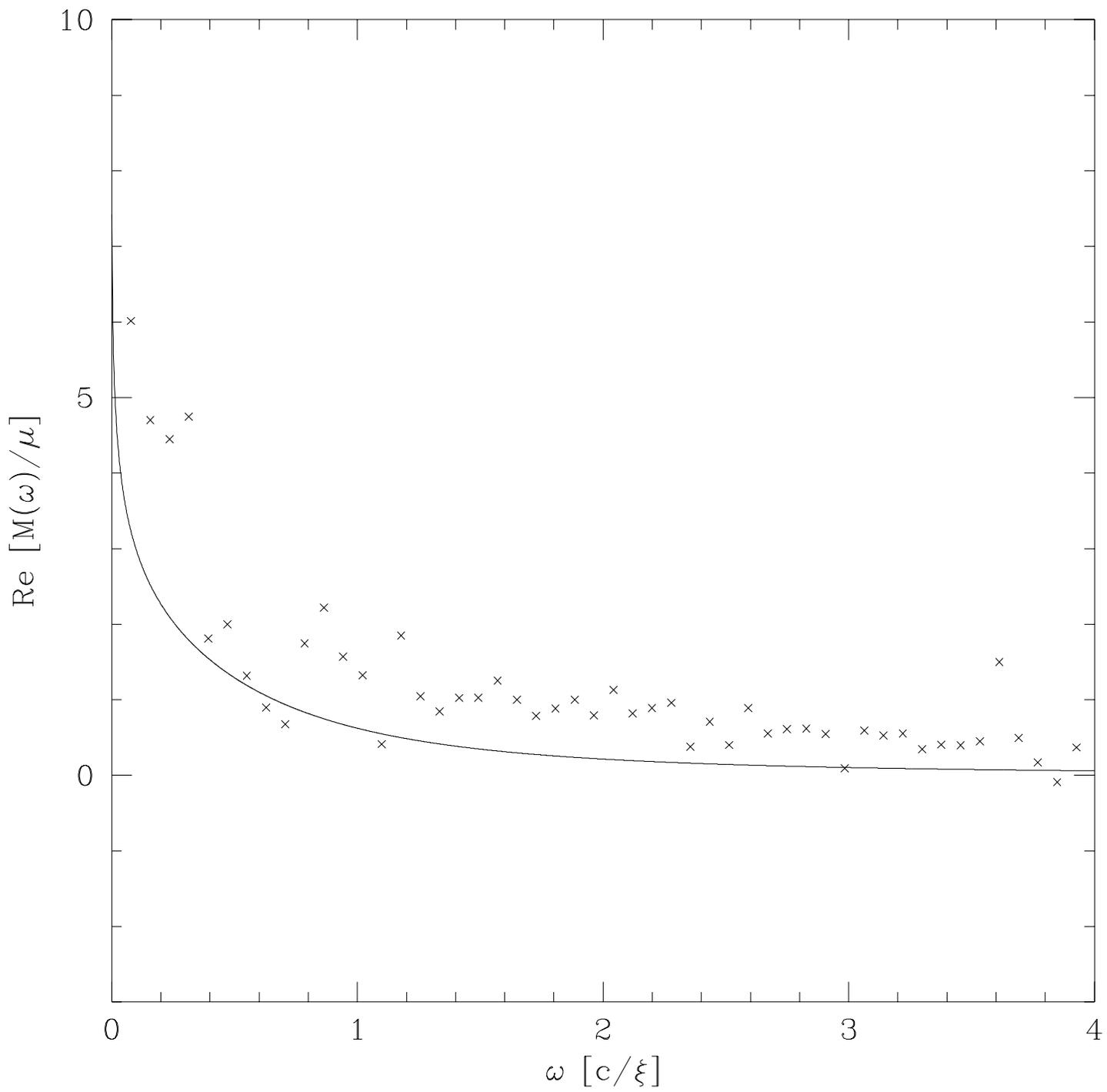
\end{document}